\documentclass[12pt,preprint]{aastex631}
  \usepackage{natbib}
\usepackage{hyperref}
\usepackage{float}
\usepackage{graphicx}
\usepackage{color}
\usepackage{amsmath}
\usepackage{ulem}
\usepackage{soul}

\begin{document}

\title{Inferring the redshift of more than 150 GRBs with a Machine Learning Ensemble model}

\author{Maria giovanna Dainotti}
\affiliation{Division of Science, National Astronomical Observatory of Japan, 2-21-1 Osawa, Mitaka, Tokyo 181-8588, Japan}
\affiliation{The Graduate University for Advanced Studies (SOKENDAI), Shonankokusaimura, Hayama, Miura District, Kanagawa 240-0115}
\affiliation{Space Science Institute, 4765 Walnut St Ste B, Boulder, CO 80301, USA.}
\affiliation{Nevada Center for Astrophysics, University of Nevada, 4505 Maryland Parkway, Las Vegas, NV 89154, USA}
\affiliation{Bay Environmental Institute, P.O. Box 25 Moffett Field, CA, California}

\author{Elias Taira}
\affiliation{Department of Physics and Astronomy, Michigan State University, East Lansing, MI 48825, USA}

\author{Eric Wang}
\affiliation{Department of Computer Science, Yale University, New Haven, CT 06511-8937}

\author{Elias Lehman}
\affiliation{Department of Physics, University of California at Berkeley, Berkeley, CA 94720, USA}

\author{Aditya Narendra*}
\affiliation{Jagiellonian University, Doctoral School of Exact and Natural Sciences, Krakow, Poland}
\affiliation{Astronomical Observatory of Jagiellonian University, Krakow, Poland}
\thanks{E-mail: aditya.narendra@doctoral.uj.edu.pl}

\author{Agnieszka Pollo}
\affiliation{National Center for Nuclear Physics (NCB), Warsaw}
\affiliation{Astronomical Observatory of Jagiellonian University, Krakow, Poland}

\author{Grzegorz M. Madejski}
\affiliation{Department of Physics and SLAC National Accelerator Laboratory, Stanford University, Stanford, CA 94305, USA}

\author{Vahe Petrosian}
\affiliation{Department of Physics and SLAC National Accelerator Laboratory, Stanford University, Stanford, CA 94305, USA}

\author{Malgorzata Bogdan}
\affiliation{Department of Mathematics, University of Wroclaw, 50-384, Poland}
\affiliation{Department of Statistics, Lund University, SE-221 00 Lund, Sweden}

\author{Apratim Dey}
\affiliation{Department of Statistics, Stanford University, Stanford, CA, 94305, USA}

\author{Shubham Bhardwaj}
\affiliation{Division of Science, National Astronomical Observatory of Japan, 2-21-1 Osawa, Mitaka, Tokyo 181-8588, Japan}
\affiliation{The Graduate University for Advanced Studies (SOKENDAI), Shonankokusaimura, Hayama, Miura District, Kanagawa 240-0115}

\date{July 2023}

\begin{abstract}

Gamma-Ray Bursts (GRBs), due to their high luminosities are detected up to redshift 10, and thus have the potential to be vital cosmological probes of early processes in the universe. 
Fulfilling this potential requires a large sample of GRBs with known redshifts, but due to observational limitations, only 11\% have known redshifts ($z$). 
There have been numerous attempts to estimate redshifts via correlation studies, most of which have led to inaccurate predictions.
To overcome this, we estimated GRB redshift via an ensemble supervised machine learning model that uses X-ray afterglows of long-duration GRBs observed by the Neil Gehrels Swift Observatory.  
The estimated redshifts are strongly correlated (a Pearson coefficient of 0.93) and have a root mean square error, namely the square root of the average squared error $\langle\Delta z^2\rangle$, of 0.46 with the observed redshifts showing the reliability of this method.
The addition of GRB afterglow parameters improves the predictions considerably by 63\% compared to previous results in peer-reviewed literature. 
Finally, we use our machine learning model to infer the redshifts of 154 GRBs, which increase the known redshifts of long GRBs with plateaus by 94\%, a significant milestone for enhancing GRB population studies that require large samples with redshift.
\end{abstract}


\section{Introduction}\label{intro}

Gamma-Ray Bursts (GRBs) are the most luminous events after the Big Bang. 
Due to their high luminosities, they are detected up to redshift $z=9.4$ \citep{Cucchiara2011}, and thus have the potential to be vital cosmological probes of processes in the early universe.
Studying GRBs enables us to deepen our knowledge about the early universe and track how the universe evolves over time since the GRB redshift range goes from 0.0085 \citep{Galma1998}) to the highest redshifts observed (between 8 and 9.4, \citep{Cucchiara2011, Tanvir2008}.
GRBs (observed mainly in $\gamma$-rays, X-rays, and sometimes in optical) are traditionally classified in short duration GRBs (SGRBs), with $T_{90}<2$s, where $T_{90}$ is the time interval during which the GRB emits 90\% of its total observed fluence (energy emitted in $\gamma$-rays) and long duration GRBs (LGRBs) where $T_{90}>2$s.
Observationally, GRBs are characterized by prompt emission, the main emission observed from hard X-rays to high-energy $\gamma$-rays and sometimes in optical \citep{Vestrand2005Natur,Beskin2010ApJ,2012MNRAS.421.1874G,2014Sci...343...38V}, and the afterglow emission \citep{costa1997,vanParadijs1997,Piro1998}, a long-lasting multi-wavelength emission, following the prompt, observed in X-rays, optical, and sometimes radio. 
The afterglow sometimes contains the plateau emission feature where the flux during the plateau remains constant \citep{Nousek2006,Rowlinson2014,Zhang2006,Dainotti2008,Sakamoto2007,OBrien2006,Zaninoni2013,Liang2007}.
Plateaus are observed in 42\% of X-ray afterglows \citep{Evans2009,Li2018b} and in 30\% of optical afterglows \citep{Vestrand2005Natur,Kann2006,Zeh2006,dainotti2020b,panaitescu2008taxonomy,panaitescu2011optical,Oates2012}.

Currently, the main issue in population studies is the lack of LGRB samples with known redshifts.
The direct determination of the redshift of a GRB requires rapid localization and spectral information. 
One of the most powerful observatories that enable rapid detection and follow-up in multiwavelengths is the Neil Gehrels Swift observatory (hereafter Swift) \citep{Gehrels2004}.
Swift uses an X-ray instrument for localization and can obtain spectra and sometimes redshift with the onboard UVOT instrument.
The Swift satellite consists of three main instruments: the Burst Alert Telescope (BAT) \citep{Burrows2005}, the X-ray Telescope (XRT) \citep{Barthelmy2005}, and the Ultraviolet/Optical Telescope (UVOT) \citep{Roming2005}. 
These instruments work together to detect, localize, and collect data on GRBs and their afterglows across various wavelengths, including $\gamma$-rays, X-rays, and ultraviolet/optical.

Swift, with its localization capabilities, has paved the way to the high-$z$ Universe. Despite all the advantages of localization provided by Swift, still, only 26.2\% (423) of Swift's 1615 GRBs have known spectroscopic redshifts up to today's date (7th of December, 2023).
Redshift measurements, particularly high-$z$ ones, are challenging due to limited telescope time and the paucity of active GRB follow-up programs.

{
{The machine learning (ML) is applied to the data, which is influenced by the observed sensitivity of the Swift satellite, e.g. flux limit, limiting energy band, occultation by the Earth (data unavailability), etc. 
The data in our sample are all taken from Swift.
Thus, the biases are all related to fluxes, the limiting energy band, and the presence of lightcurve gaps in relation to data observed by the same satellite. 
Therefore, we have a uniform bias, and since our generalization sample is also taken entirely from the Swift satellite, the impact of this bias is reflected in the same way both in the training set and in the generalization sample. 
This means that the impact of this bias in the prediction is similar to the biases we have in the observed sample.
If one would like to apply a bias correction, for example, consider the Malmquist bias effect that allows only the brightest GRBs, with larger fluxes, to be observed, then one needs to employ the redshift.
However, we cannot use the information of the redshift in training the ML models since the redshift is the variable we would need to determine. 
This, indeed, would induce a circularity argument.
Thus, beyond the analysis of a posteriori bias correction, we cannot apply any further correction to the initial data.
If we had applied any such treatment to the data, we would have induced a data bleeding problem in our training set. 
So, in summary, the impact on the training sample is negligible, considering that here we use a uniform data set both in the training and the generalization.}
}

Thus, efforts to determine the redshift of GRBs are of paramount importance.
One of the key goals of increasing the sample of GRBs with known redshift is to determine an accurate measure of the luminosity function (LF), which provides the number of bursts per unit luminosity, key to understanding the properties of GRB luminosities as a population, the energy release and emission mechanism of GRBs. 
Another relevant goal is the determination of the cosmic GRB formation rate (GRBFR), which provides the number of events per comoving volume and time, pivotal to understanding the production of GRBs at various stages of the universe.
As highlighted in \cite{Petrosian2015}, there exists a discrepancy between LGRBs compared to the overall rates of star formation within the lower redshift range ($0 < z < 1$). 
This result has been found by several groups with differences, and thus, the debate is still open. Obtaining more redshifts becomes crucial for settling such a debate.

Another great advantage of having more GRBs with redshift is the possibility of using GRBs as standardized candles with empirical relations between distance-dependent and intrinsic properties of GRBs.
Amongst the earliest of these efforts, is the Dainotti Relation  \citep{Dainotti2008,Dainotti2011a,2015ApJ...800...31D,Dainotti2017}, a roughly inversely proportional relationship between the rest-frame time at the end of the plateau phase ($T_{a}/(1+z)$) and its corresponding luminosity ($L_{a}$).
Later \cite{Dainotti2013} showed that via the use of the Efron and Petrosian method (\cite{EffronPetrosian1992}), this relation is intrinsic and not due to selection biases. 
This relation has also been discovered in the optical and radio emission \citep{dainotti2020b,Levine2022}.
It has also been extended in three dimensions in X-rays, where the peak prompt luminosity $L_{peak}$ has been added to the two-dimensional Dainotti relation \citep{Dainotti2016,dainotti2017b,Dainotti2020}. 
In addition, GRBs observed by Fermi-LAT and detailed in the Second Fermi GRB Catalog \citep{2019ApJ...878...52A}, which show the existence of the plateau in $\gamma$-rays, obey this correlation as well \citep{2021ApJS..255...13D}. 
Continuing on the extension of this relation in other wavelengths, this three-dimensional relation has been found in optical too \citep{Dainotti2022c}.

Both the two and three-dimensional relations have been used as a valuable cosmological tool \citep{2009MNRAS.400..775C, 2010MNRAS.408.1181C, 2013ApJ...774..157D, 2014ApJ...783..126P,Cao2021,CaoShulei2022,CaoShulei2022a,Dainotti2022b, Dainotti2023a, Bargiacchi2023, Dainotti2023b}.
\cite{Dainotti2022b}, showed how the Dainotti relation used in combination with SNe Ia is able to obtain consistent results on matter density, $\Omega_M$ to SNe Ia in the $\Lambda$CDM model with the added benefit of extending the distance ladder up to $z = 5$, a redshift far greater than the farthest observed SNe Ia observed up until $z=2.26$ \citep{rodeny2015}. 
Another large part of this paper discussed the prediction of the number of GRB observations that we would need to obtain the same precision as SNe Ia on the matter density, $\Omega_M$ in the $\Lambda$CDM model as in \cite{Conley2011,Betoule2014,Scolnic2018}. 

Indeed, it has been discussed in \cite{Dainotti2022a} that we need 789 GRBs to reach the same precision of SNe Ia in {\cite{Conley2011}. 
Thus, we need to add 567 more GRBs to our current sample of 222 GRBs with X-ray plateaus and known redshift. 

To increase the number of GRBs with redshifts, there have been several attempts in this direction by finding correlations between distance-independent quantities (peak flux, duration of afterglow plateau, as the already mentioned \cite{Dainotti2008, Dainotti2013}, etc) and distance-dependent GRB properties (prompt emission peak luminosity) to find pseudo-redshifts for the GRBs with unknown redshift (\cite{Atteia2003}, \cite{yonetoku2004}, \cite{Dainotti2011}). 
The results of these analyses are all reliant on the luminosity distance ($D_L$), a quantity that, by definition, not only depends on cosmology but small variations of the $D_L$ at high redshift are subject to large variations of the redshift. 
Thus, these results are inherently subject to inaccuracy. 
To avoid the issues caused by including $D_L$ in determining pseudo-redshifts, we undertake a new approach that relies on the use of ML algorithms to create our redshift predictions.

The paper is structured as follows: in Sec. \ref{intro} we detail the problem of paucity of the redshifts, in Sec. \ref{datasample}
we describe the dataset. 
In Sec. \ref{methodology}, we describe our pipeline from how we process our data to select the variables to be used to how we build and test our models on our data. In Sec. \ref{results} we discuss the performance of our model as well the predictions on the generalization set and we compare those results to the distribution of the existing set of redshifts. Finally in Sec. \ref{discussion} we summarize and discuss the implications of these results.}

\section{The Data Sample}\label{datasample}

In this study, we focus on GRBs observed in $\gamma$-rays and X-rays detected by the BAT and the XRT telescopes onboard the Swift.
We used the data stored in the \href{https://swift.gsfc.nasa.gov/archive/grb_table/}{NASA Swift GRB Search Tool}, and the Third Swift-BAT GRB Catalogue \citep{Lien2016}.
Our initial step involves preprocessing the raw GRB data to ensure its quality and suitability for further analysis. 
Considering the distinct nature of various GRB classes, such as LGRBs and SGRBs, which can originate from different progenitors or the same progenitors in diverse environments, it is crucial to avoid blending the characteristics of these diverse classes.
For this reason, our study focuses only on LGRBs.
Thus, we exclude from our sample, taken from \cite{Dainotti2020} and \cite{Srinivasaragavan2020}, SGRBs, SGRBs with extended emission \citep{Norris2006} and the intrinsically SGRBs (IS) which have $T_{90}/(1+z)<2$s.
The initially available features are the following:

\begin{enumerate}
\item $T_{90}$ - the time interval during which the GRB emits 90\% of its total observed fluence (energy emitted in $\gamma$-rays).
\item $F_a$ - the flux at the end of the plateau emission.
\item $T_a$ -  the time at the end of the plateau emission.
\item $\alpha$ - the temporal power-law index 
after the end of the plateau emission.
\item $\beta$ - the spectral index assuming a power-law for the spectral energy distribution in the range of the plateau emission.
\item $\gamma$ - the spectral index {obtained as the time-averaged spectral fit from the Swift XRT Photon Counting mode data.}
\item Fluence - the energy fluence over $T_{90}$ of the prompt emission in units of erg {cm}$^{-2}$.
\item PhotonIndex - the prompt photon index {from the BAT Telescope} of the photon energy distribution modeled with a power law.
\item NH - The column density of neutral hydrogen along the line of sight.
\item Peak Flux - The prompt peak photon flux in units of  with unit of (number of photons) cm$^{-2}$ s$^{-1}$.
\end{enumerate}

The dataset with which we train and test our ML models contains 197 LGRBs with all the features listed above, called the predictors, as well as our response variable, the redshift. 
Furthermore, there are also 221 GRBs without a measured redshift and the same features. 
This set is called the generalization set. The ML models are used to predict the redshift of these GRBs.

\section{Methodology}\label{methodology}

\begin{figure}[!h]
    \centering
    \includegraphics[scale=0.8]{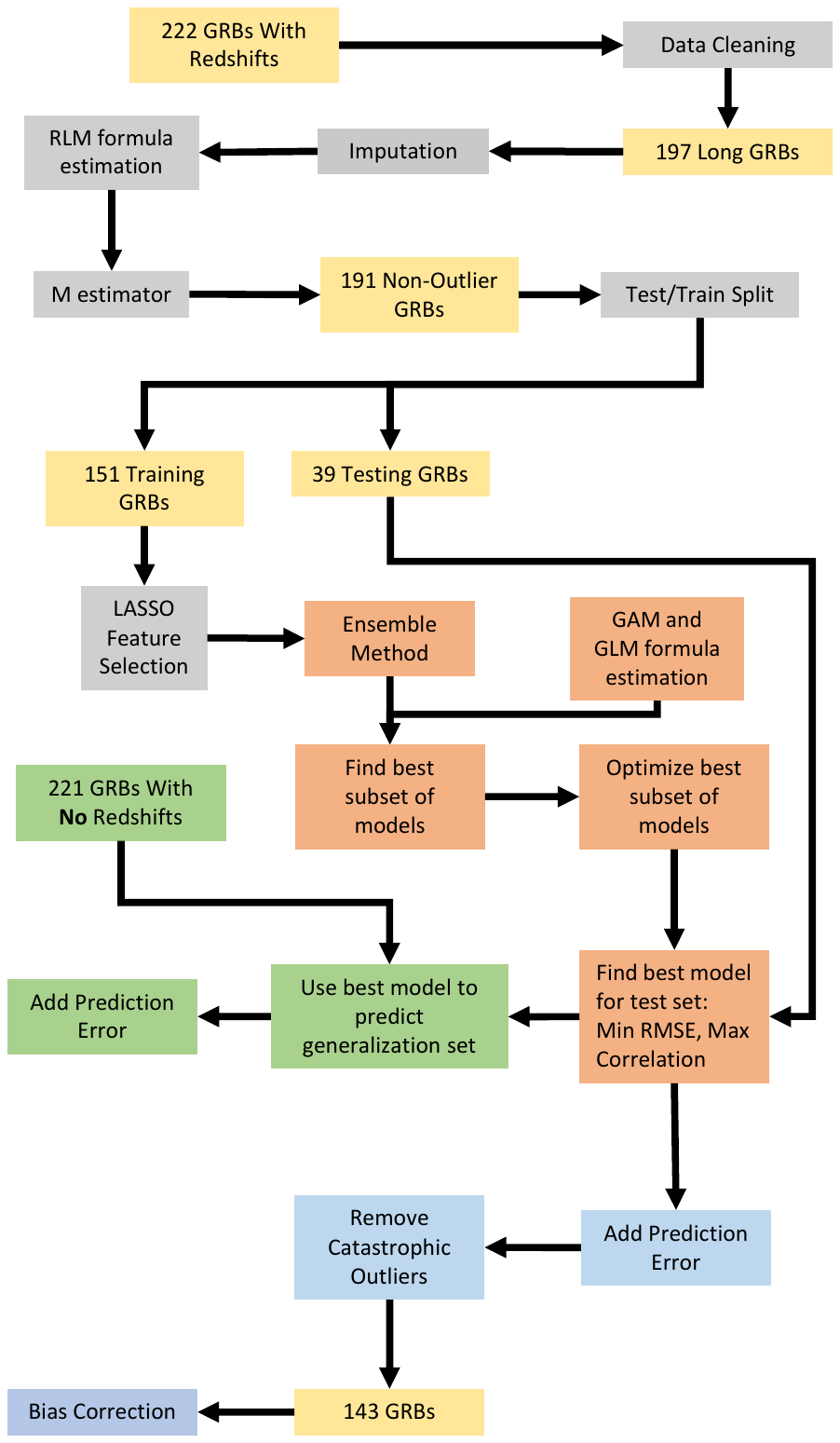}
     \caption{The flowchart detailing each step of the pipeline. All yellow boxes represent indicate the number of GRBs in the training set at each step of the process. Grey boxes are pre-processing steps. Orange boxes indicate all steps involving model construction. Green boxes show all steps involving the generalization set predictions. Blue boxes show all the post-processing steps with the training set.}
    \label{fig:flowchart}
\end{figure}

Fig. \ref{fig:flowchart} shows the summarized flowchart for our methodology. 
The individual parts are expanded upon in the following subsections.

\subsection{Data Cleaning and Transformation}\label{sec:datacleaning}

\begin{figure}[!h]
    \centering
    \includegraphics[scale=0.3]
    {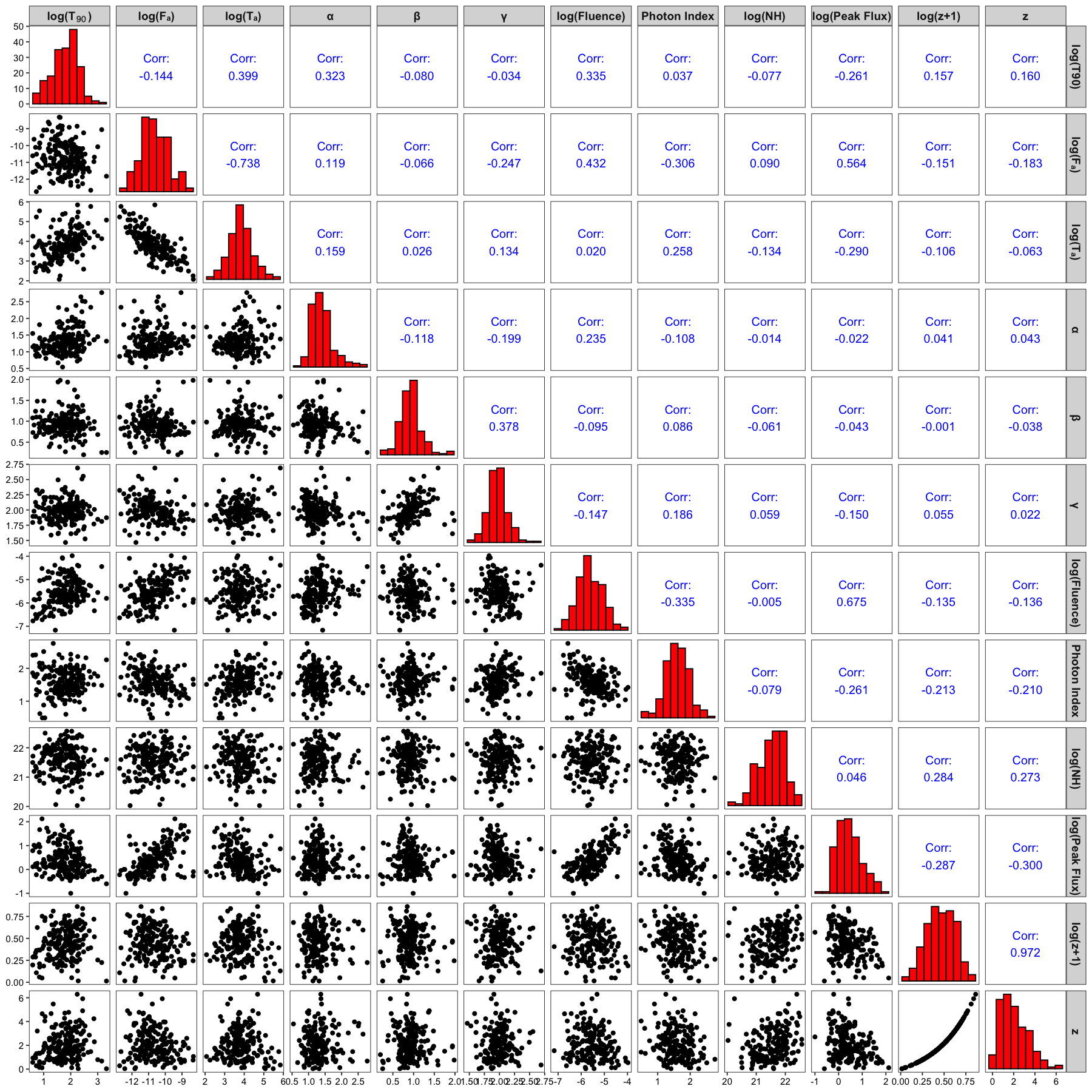}
    \caption{Scatter matrix plot of the data after the cleaning of the sample and variable transformation.}
    \label{fig:init_scatter}
\end{figure}

Due to the wide range encompassed by specific variables, namely Peak Flux, $T_a$, Fluence, NH, $T_{90}$, and $F_a$, we transformed these variables into log base-10 with the aim of enhancing prediction accuracy. 
The variables, $\alpha$, $\beta$, $\gamma$, and PhotonIndex remain in the linear scale since the range in which they vary is of the order of unity. 
{We then proceed to clean our data such that we exclude any non-physical values from our analysis or values which are unusual for the majority of GRBs such as $\alpha > 3$ {(2.5\%)}, $\beta > 3$ {(1\%)}, $\gamma >3$ {(0.5\%)}, Photon Index $< 0$ {(0.5\%)} { and $\log(NH) < 20$ (8.6\%)} so that we can capture the average features of the GRBs.} 
{All these values are set to NA and then imputed (see Sec. \ref{MICE}).}
{We exclude the values of $\alpha$, $\beta$, and $\gamma$ $>$3, because these belong to the tail of their respective distributions, as can be seen in Fig. \ref{fig:outlierhist}}.
We also perform a similar transformation to our redshifts in which we create the new response variable $\log(1+z)$. 
This results in a Gaussian distribution with a mean=0.48 and a standard deviation=0.128 for this response variable, rather than a distribution with tails as shown for the distribution of the redshift in the scatter matrix plot of Fig. \ref{fig:init_scatter}.
This new response variable is chosen similarly as in previous literature (\cite{Dainotti2021}, \cite{Gibson2022}, \cite{Narendra2022}), and it is a natural choice since it mimics the evolution of the variables.
In addition, z+1 is a more natural parametrization of the cosmological variable z.
We show the scatter matrix plot after the data cleaning and transformation, see Fig. \ref{fig:init_scatter}.

\subsection{Data Imputation: MICE}\label{MICE}

Multivariate Imputation by Chained Equations (MICE, \cite{van2011mice}) can impute missing values for multiple variables using variables from the data set that are complete. 
MICE has the ability to create imputed values in R with a variety of different methods. 
Here, we use the predictive mean-matching method known as ``midastouch" to create our model.
{We employ the ``midastouch" approach, a predictive mean matching (PMM) technique introduced by \cite{little2019statistical}. 
This method begins by populating missing values in a feature with its mean and subsequently estimating these values by training a model on the available complete data. 
For each prediction, a probability is assigned based on its distance from the value imputed for the missing variable.
The missing entry is imputed by randomly drawing from the observed values of the respective predictor, weighted according to the probability defined previously.}

This process is then repeated $N$ times, after which the final substituted quantity for each missing value is determined by taking an average over the prediction of the value in each iteration. 

{In previous literature \citep{Gibson2022}, similar methodologies have been applied to Active Galactic Nuclei data from the Fermi Fourth LAT catalog with no noticeable addition to the uncertainty of the resulting data distribution. 
In fact, the constructed ML model strictly benefits from its application due to the increased size of the dataset. 
Given that GRBs also exhibit similarly nonlinear trends within their features, we expect to see similar results in our own study.}

\begin{figure}[!h]
    \centering
    \includegraphics[scale=0.3]
    {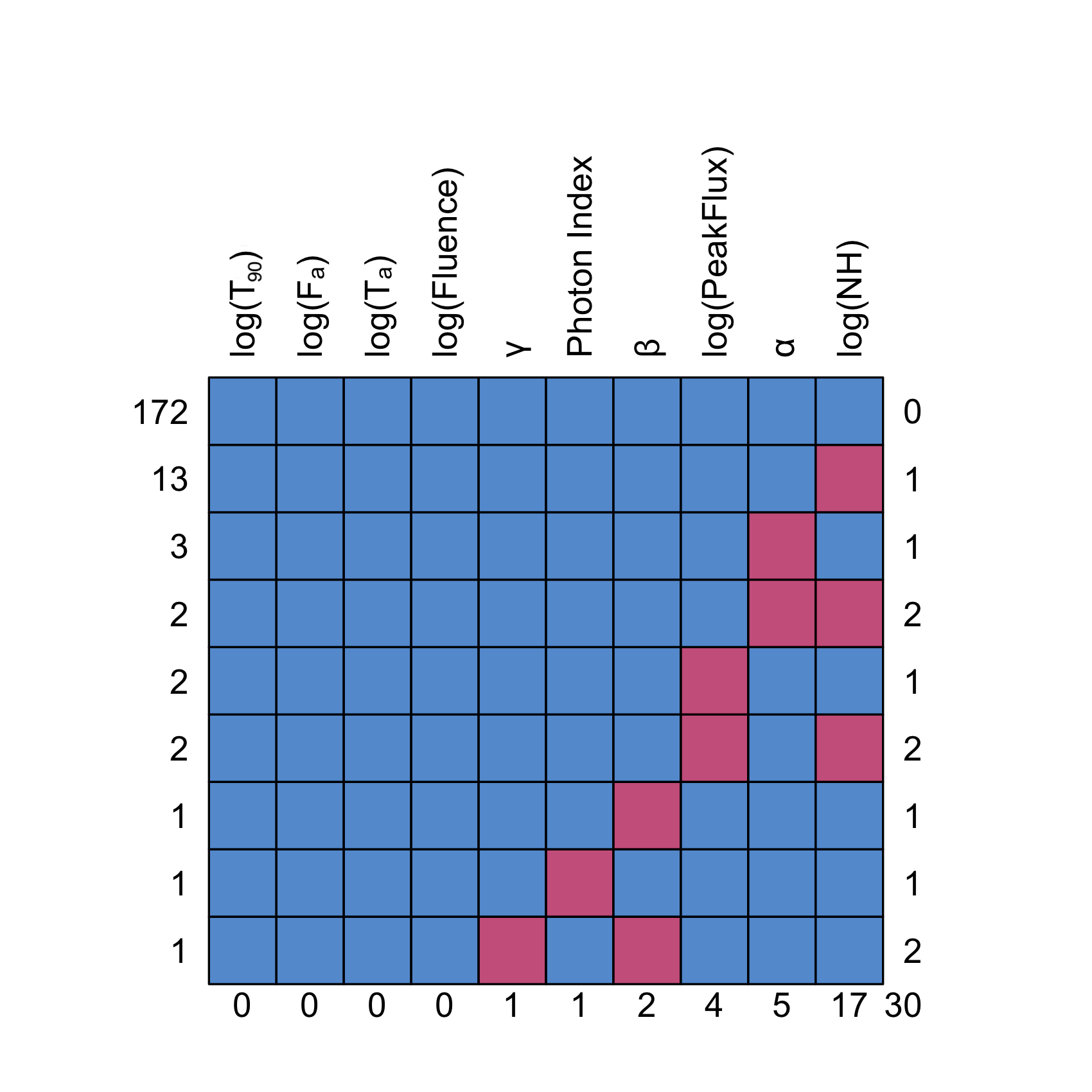}
    \caption{The missing data in our sample. The red boxes show the missing GRB data points, while the blue boxes indicate GRBs with no missing data for a given GRB variable presented in the top axis. The bottom axis shows the number of missing GRBs per variable. The left axis represents the number of observations that have missing data for a specific set of features.
    For example: there are 172 GRBs with no missing data, 13 GRBs with missing data in $\log(\text{NH})$ data, 3 GRBs with missing data in $\alpha$, and so on. The right axis represents the number of features that are missing for that row.}
    
    
    \label{fig:mice}
\end{figure}

Here, we show the missing data in Fig. \ref{fig:mice}. 
The bottom x-axis shows the number of missing GRBs corresponding to the variable presented in the upper x-axis. 
The pink boxes show the missing GRB variables, while the blue boxes indicate GRBs with no missing data for given variables. 
We now detail the missing data points in our variables: 
{1 GRB has missing data in $\gamma$, 1 has missing data in PhotonIndex, 2 have missing data in $\beta$, 4 have missing data in $\log(\text{Peak Flux})$, 5 have missing data in $\alpha$, and 17 have missing data in $\log(\text{NH})$}.
While $\log(T_{90})$, $\log(F_a)$, $\log(T_a)$, and $\log(\text{Fluence})$ have no missing data points.

\subsubsection{Nested 10-fold Cross Validation and the Extensive Search} \label{ExtSearch}
\label{10fCV}
Here, we describe a nested 100-iteration 10-fold cross-validation (10fCV) procedure that will be utilized in both the outlier removal and model construction stages. 
This procedure is called nested since it requires an external layer of cross-validation.
10fCV involves dividing our data set into 10 distinct subsets, each containing 10 parts.
We then iteratively train the model using 9 out of the 10 subsets as training data and evaluate its performance on the remaining subset as a testing set. 
This procedure is repeated for each subset, allowing each subset to serve as a testing set, while the others are used for training.
We average the prediction results across all the 100 iterations to obtain the mean prediction. The standard deviation of this distribution is the prediction error.

The extensive search we perform uses the nested 100-iteration 10fCV. 
The procedure preparatory to the extensive search requires first building all possible formula candidates. 
These formulas are meant for two purposes. 
The first purpose is generating multiple models for the robust linear model (RLM) using the M-estimator procedure, as explained in Sec. \ref{OutlierRemoval}. 
The second purpose is to create multiple models for both the Generalized Additive Model (GAM) (see in Sec. \ref{GAM}) and the Generalized Linear Model (GLM) (see in Sec. \ref{GLM}), both collectively used to construct the final ensemble method.
We build the formulas described above with a generator function that employs the first-order features, which are the observed GRB variables, and the second-order variables, which are the multiplicative terms among the first-order variables.
Then, the extensive search allows us to find the best formula among all formulas tested based on the correlation between the predicted and observed redshift and the root mean square error (RMSE), a measure of deviation from the model's fit written for $N$ data points as

\begin{equation}
    \text{RMSE}=\sqrt{\frac{\sum_{i=1}^N(x_i-\hat{x}_i)^2}{N},}
\end{equation}
where  $x_i$ is the true response value for data point $i$, and $\hat{x}_i$ is the respective predicted value. 

For GAM, RLM, and GLM, each has its own individual cut-off for the correlation and RMSE. This cut allows us to choose the best formulas.
Following this, we use these formulas to predict the redshift of the test set.
Next, we identify the formula corresponding to the highest correlation, the lowest RMSE, and the lowest median absolute deviation (MAD), which is the median of the absolute difference between each data point and the mean of the dataset given by

\begin{equation}
    \text{MAD}=Median(|x_i - \Bar{x}|),
    \label{MAD-EQ}
\end{equation}

where $x_i$ are the data points and $\Bar{x}$ is the mean of the dataset $x$.
Finally, we select the best formula from these based on the highest weight in the SuperLearner (detailed in Sec \ref{SL})

\subsubsection{Outlier Removal}
\label{OutlierRemoval}
In order to remove outliers we use a preliminary robust regression method, M-estimation, which minimizes the residuals in a given model.
The application of the M-estimator enables the fit of an RLM on the imputed data. 
We conduct an extensive search (see Sec. \ref{10fCV}) to find an optimal formula for the RLM that best fits the data.


To this end, we include square terms of one or multiple features of the data to capture potential non-linear relationships between our predictors and the response variable. 
The chosen model reads as follows: 

\begin{equation}
\begin{aligned}
\log(1 + z) = (&\log(\text{NH})^2 + \log(T_{90})^2 + \log(T_a)^2 + \log(\text{NH}) \\
&+ \text{PhotonIndex} + \log(T_{90}) + \log(T_a) + \log(F_a))^2 \\
&+ \log(\text{PeakFlux}) + \log(F_a)^2 + \text{PhotonIndex}^2 + \log(\text{PeakFlux})^2.
\end{aligned}
\label{RLM}
\end{equation}

M-estimator is an alternative technique to the ordinary least squares method, which fits the function mentioned above to our data.
The ordinary least squares method attempts to minimize the square of the residuals (called the $L_2$ norm regression) by giving outliers of the data set a higher weight. This significantly affects the results of the regression fit.
In contrast, the M-estimator attempts to minimize the sum of a function of residuals. 
The function chosen for our analysis is the Huber Function \citep{huber1964}. 
RLM is used for the detection of highly influential observations. 
We are using the implementation of RLM as described in the MASS package of R \citep{venables2002random}.
Data points with weights falling below 0.5 undergo exclusion, a crucial step taken to counterbalance the influence that potentially problematic data points may exert on the model's effectiveness. 
Following this outlier removal procedure, we eliminate 6 outliers (GRB050826, GRB051109B, GRB080916A, GRB111008A, GRB151112A, and GRB160327A), reducing the size of our data set to 191 GRBs.

\subsubsection{Feature Selection}
The preprocessed data is now divided into two sets: an 80\% training set for model training and a 20\% test set for performance evaluation which is never used for the best model selection.
We decide to reduce the number of variables to be investigated, and thus we select the most predictive features given the small data sample.
To identify the most important features, we use the Least Absolute Shrinkage and Selection Operator (LASSO) method exclusively on the training set \citep{Tibshirani1996}. 
To ensure the stability of the results, we perform LASSO for 100 iterations and obtain as a result the averaged weights for each predictor.
{To reduce the number of features we chose only those that have a non-zero LASSO weight.
We extract the following features:} $\log(T_{90})$, $\log(F_a)$, $\log(T_a)$, PhotonIndex, $\log(\text{NH})$, and $\log( \text{Peak Flux})$.
These features will be used in all the successive steps of our analysis to find the most predictive model.

\begin{figure}[!h]
    \centering
    \includegraphics[scale=0.35]{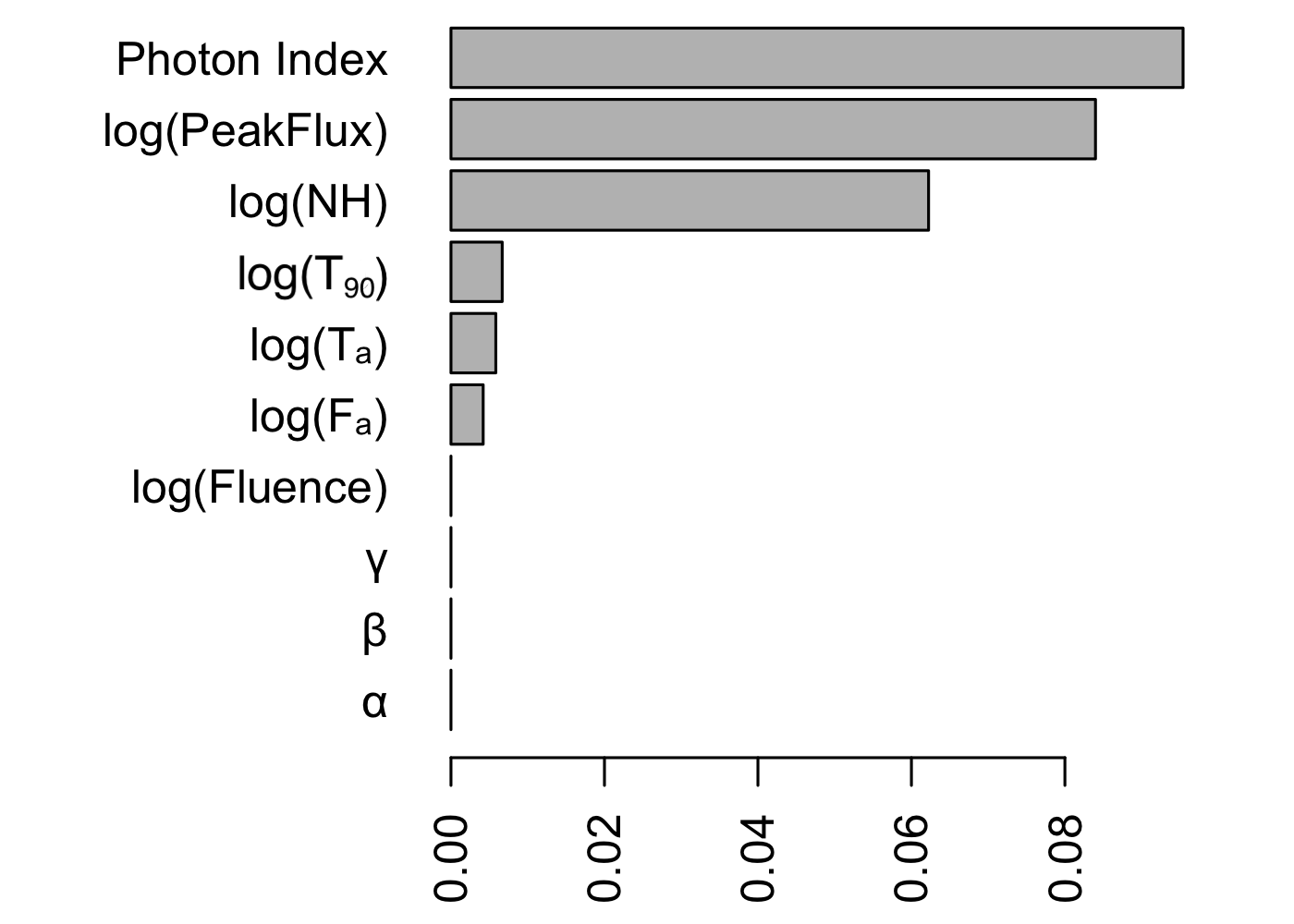}
    \caption{The weights assigned to the features by LASSO.}
    \label{fig:LASSO}
\end{figure}

\subsection{Model Construction}
\label{Model Construction}

Once the data has been preprocessed, we begin constructing an ensemble of supervised ML algorithms to model the relationships between the GRB features and their corresponding redshifts.
Supervised ML leverages prior knowledge of the ‘training’ data on which {ML models will be built, and their predictions will be tested} on new data (the ‘test’ set).
{Parametric models} use functions with a set of parameters whose coefficients are fine-tuned to fit the training data. 
These models, although simpler and faster to train, are however constrained by the functions.
Non-parametric models, in contrast, without assuming a predefined function, are thus more flexible and powerful than parametric models. However, they are prone to over-fitting, require large samples, and extensive running time. 
Semi-parametric models have a parametric and a non-parametric component, thus leveraging the advantages of both models. 

We begin our construction by testing a total of 115 different regression methods on 10 iterations of 10fCV. 
These 115 models include the best of the following ML algorithms, which were tested individually:
Random forest models, with differing numbers of trees, ranging from 10 to 500; 
extreme gradient boosting models using the same tree combinations.
Further, 92 models from the caret package \citep{caret} were also tested.
And finally, a single support vector machine (SVM) model was picked from a combination of 7921 models. 
These combinations were obtained by changing the various hyperparameters of the SVM model.



Out of these 115 models 25 models were selected for obtaining the highest correlation during the surveys.
Namely, these were GAM; GLM, Bayesian GLM, GLM Network, and Interaction GLM; Extreme Gradient Boosted trees; Recursive Partitioning and Regression Trees (RPart) and Random Forest as implemented by the caret package; Random Forest with Conditional Inference, R's native RPart, RPart with Pruning, Bagged Trees, Fast Implementation of Random Forest (ranger); Ridge; Stepwise Akaike Information Criterion, Interaction Stepwise Regression, Forward Stepwise Regression, and Classical Stepwise Regression; Feed Forward Neural Network; Regression Towards the Mean; Local Weigthed Regression; Linear Modeling; a tuned Kernel Support Vector Machine; Fast Multivariate Adaptive Regression Splines; and a scalable version of Lasso.

Our search for the best ML models consistently exhibited a preference for linear parametric and semi-parametric models. 
The resilience of these models leads us to believe that standardizing a tuning methodology for the non-parametric models is not required for this work. 
As a general remark, we would like to stress that, in principle, the fully non-parametric models have the advantage of estimating complicated non-linear relationships between the response variable and the predictors as well as high-order interactions between features. 
Such non-parametric methods are very powerful when the data set contains many observations. However, they suffer from the so-called “curse of dimensionality”, which sets limits on the number of parameters one can efficiently estimate for a given sample size. 
This “curse” becomes more severe with a greater number of features. 
As a result, the fully non-parametric ML methods allow the use of only a limited number of features, e.g. when estimating redshifts based on small GRB training sets, as in this case.



\subsubsection{The Generalized Linear Model}
\label{GLM}

GLM is a parametric regression technique that utilizes specialized link functions to relate the distribution of the response variable to a linear combination of the predictors. 
As opposed to the standard linear model, GLM excels in its ability to handle various distributions, such as Gaussian, Poisson, and Gamma, by selecting the appropriate link function. 
The model's parameters are estimated through Maximum Likelihood Estimation (MLE), iteratively refined for optimal fit \citep{Nelder1972}.  
This allows us to explore different model architectures and identify the most suitable model for our data. 
We perform the extensive search over 4158 formulae (see the left panel of Fig \ref{fig:final_model_scatter_plot}), composed of first and second-order variables. 
{The cutoffs shown in the right panel Fig 5 include only the formulae with $r$ above the 99.5th percentile ($r > 0.587$) and $RMSE$ below the 10th percentile ($RMSE < 0.158$).}
The formula that obtains the best correlation on the test set (see Sec \ref{ExtSearch}) is selected as the final formula.
In our implementation of GLM, we assign the formula below as the desired fitting function. 
This is based on our results from the extensive search (see Sec. \ref{ExtSearch}) with a Gaussian link function:


\begin{equation}
\begin{aligned}
\log(1 + z) = &(\log(NH)^2 + \log(T_{90})^2 + \log(T_a)^2 + \log(NH) \\
&+ \text{PhotonIndex} + \log(T_{90}) + \log(T_a) + \log(F_a))^2\\\
&+ \log(\text{PeakFlux}) + \log(F_a)^2 + \text{PhotonIndex}^2 + \log(\text{PeakFlux})^2.
\end{aligned}
\label{eq:GLM}
\end{equation}
Note that this formula is same as the formula obtained for the outlier removal (Eq. \ref{RLM}).
\newpage
\subsubsection{The Generalized Additive Model}\label{GAM}

        
In the semi-parametric GAM \citep{Hastie1990}, the redshift is related to the GRB variables via the sum of either parametric or non-parametric functions including smooth functions or a combination of both.
The advantage of GAM is that it incorporates smooth functions on specified features to relate nonlinear relationships between the features and the response variable. 
Each of the smooth functions is represented using a group of basis functions, also known as B-splines, which compose a piecewise polynomial function to relate the smoothed predictor to the response, constrained by a specified degree of freedom. 
In our implementation of GAM, however, a penalty term is applied to the B-splines to penalize high complexity, eliminating the need for manually specifying the degrees of freedom. 

\begin{figure}
    \centering
    \includegraphics[scale=0.5]
    {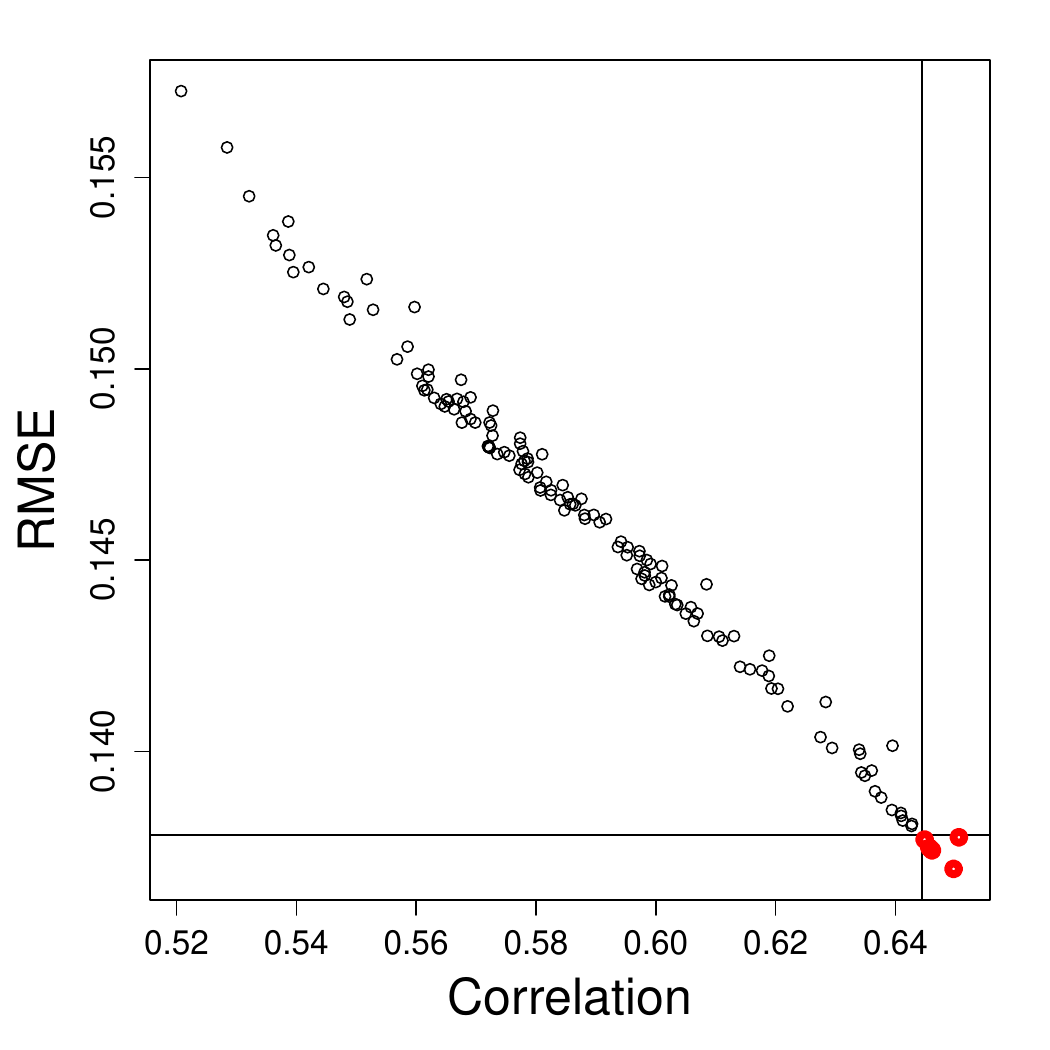}
    \includegraphics[scale=0.5]
    {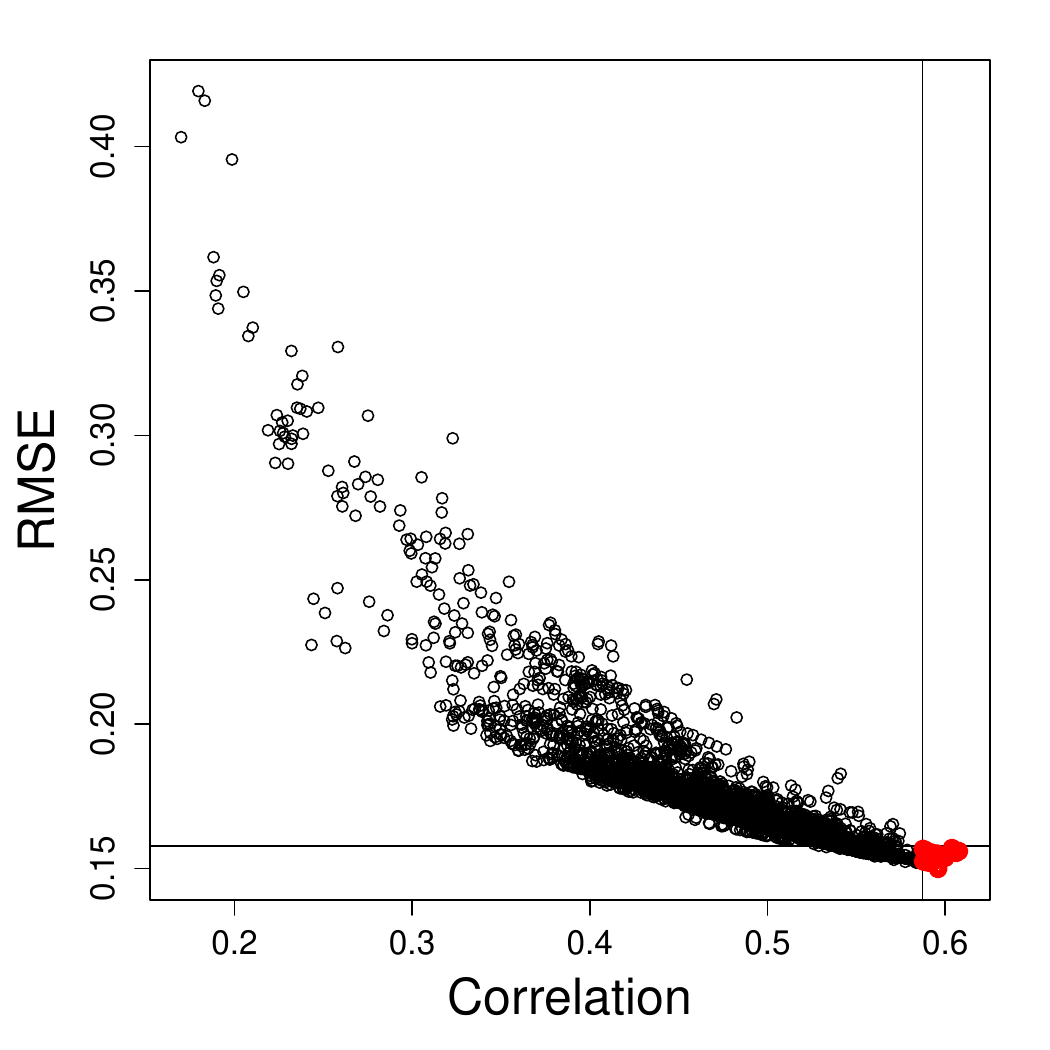}
    \caption{
    The plot of the cross-validation results of RMSE and the $r$ in the $\log(z+1)$ of GLM (right) and GAM (left) formulae.
    Each dot on either plot represents a formula performance within the 10fCV. The red dots represent the formulae that were above the chosen RMSE and Correlation.
    }
        \label{fig:final_model_scatter_plot}
\end{figure}

Here, we emphasize the significance of utilizing first-order variables, as opposed to second-order variables. 
The use of second-order variables possesses the following disadvantages: firstly, the error related to the features would be squared and further propagated onto the redshift. 
In addition, there would be unnecessary complexity added to the model, especially considering that formulas with first-order variables with similar prediction quality exist. 
For these reasons, by prioritizing first-order variables, we enhance the precision of our predictions and subsequent error estimation.
Thus, our extensive search for GAM is conducted on 141 formulae (see the right panel of Fig \ref{fig:final_model_scatter_plot}), composed of both smoothed and unsmoothed first-order variables. 
{We employ cuts to include only the formulae with $r$ above the 97th percentile ($r > 0.644$) and $RMSE$ below the 3rd percentile ($RMSE < 0.138$) for the final selection based on performance on the test set (see Sec \ref{ExtSearch}).
These cutoffs are different for GLM and GAM because we are conducting the two searches over different numbers of formulae: 4,158 for GLM and 141 for GAM. 
Therefore, it is appropriate to apply different quantile cutoffs.}
We arrive at the below-mentioned formula for GAM:
\begin{equation}
\begin{aligned}
    \log{(1+z)} = &s(\log{\text{(NH)}}) + s(\log{(T_{90})}) + s(\log{(T_a)}) \\ 
    &+ \log{(F_a)} + \text{PhotonIndex} + \log{\text{(Peak Flux)}}.
    \label{eq:GAM}
\end{aligned}
\end{equation}
{Here \textit{s()} denotes the smoothing function applied to the parameters in the GAM formula described above.}



\subsubsection{SuperLearner}
\label{SL}
The SuperLearner is an ensemble of ML models that has the advantage of combining several ML methods into a single model and leveraging the predictive power of each singular model. 
It is also able to use the same model with varying configurations and assess how each of these models performs.
However, to perform this assessment, the models must be constructed in such a way as to minimize the anticipated risk, which quantifies the model accuracy, by reducing the RMSE.

Outside of the algorithm, we employ a nested 10fCV 100 times (as mentioned in Sec \ref{10fCV}) to gauge the accuracy of each individual ML model.
By analyzing the ensemble's behavior across different runs, we aim to identify the best-performing set of models and their corresponding weights.
Subsequently, SuperLearner creates an optimal weighted combination of these models, providing an ensemble based on the performance on the test data. 
SuperLearner provides coefficients that indicate the weight ($A_i$) or significance of each individual learner within the collective ensemble. 
By default, these weights are non-negative and sum to 1.
This approach has been demonstrated to achieve asymptotically the same accuracy as the most effective prediction algorithm in the ensemble.

We survey 115 models individually, as described in Sec. \ref{Model Construction}, and we employ 25.
Then, we use weights assigned by SuperLearner as a discriminator among the most predictive models, and we remove models weighted $<0.25$ after having performed a 100 10fCV to ensure the stability of the most performative models.

\begin{figure}
    \centering
    \includegraphics[scale=0.5]{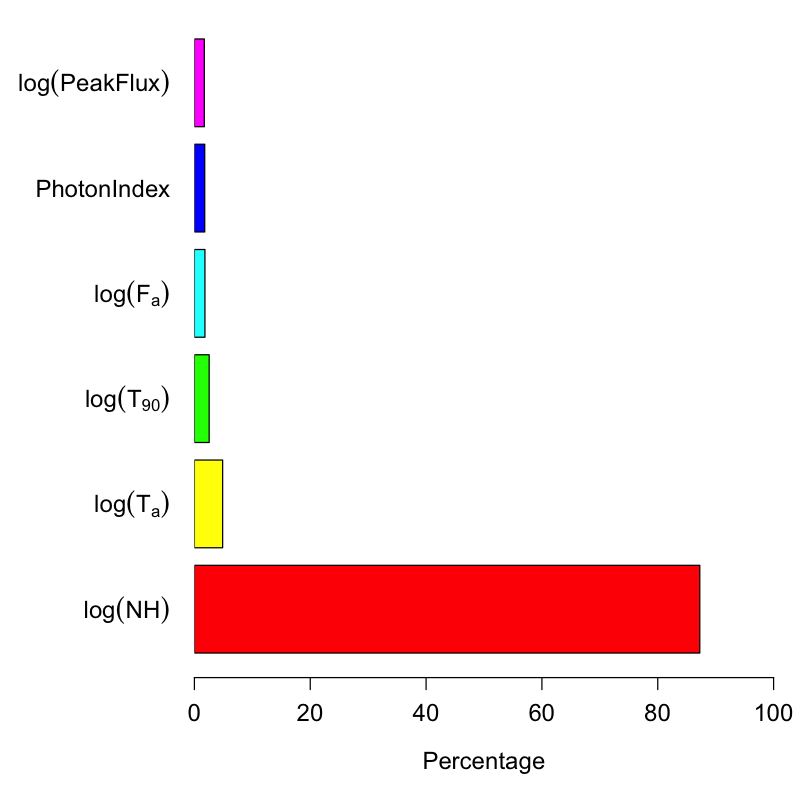}
       \caption{Plot of relative influence of each predictor including in each predictor also the second-order variables.}
    \label{relativeinfluence}
\end{figure}

\subsection{The relative importance}

To assess the contribution of each predictor, we use the relative importance of features which is an average of local linear approximation of prediction. 
For each observation, synthetic data is generated by adding Gaussian noise.
Subsequently, we construct an approximate change in prediction through a linear model prediction, $P= X \times B$ where B are fit coefficients.
The local relative importance of feature $i$ is defined by $R_i = |Bi|/P|Bi|$ for all sample points.
In Fig. \ref{relativeinfluence}, the bars show the relative influence of the variables, where each bar contains the sum of the relative influence of the first-order and second-order variables. 
As we can see, the second most important variable is $T_a$, highlighting the importance of adding the plateau emission among the features. 
Indeed, all the variables related to the plateau emission survive the trimming performed by LASSO, and these variables, $T_a$, Peak Flux, and $F_a$, naturally recover the Dainotti correlation discussed in the Sec. \ref{intro}.



\subsubsection{Error prediction and the Catastrophic Outlier Removal} \label{errorbars}

The GRBs in our sample have error measurements for $\log(F_a)$, $\log(T_a)$, $\log(T_{90})$, $\log($Peak Flux) and PhotonIndex. 
The uncertainties on $\log(NH)$ are difficult to gather since they are not present in the BAT catalog and for many GRBs such uncertainties are lacking.
Thus, in the measurement errors we do not include the $\log(NH)$ uncertainties.
In order to account for these observational uncertainties, we perform a Monte-Carlo Markov Chain (MCMC) approach, making the assumption that our uncertainties are Gaussian, because the uncertainties on the variables are independent between measurements of the same variable, and are random.

\begin{figure}
    \centering
    \includegraphics[scale=0.27]{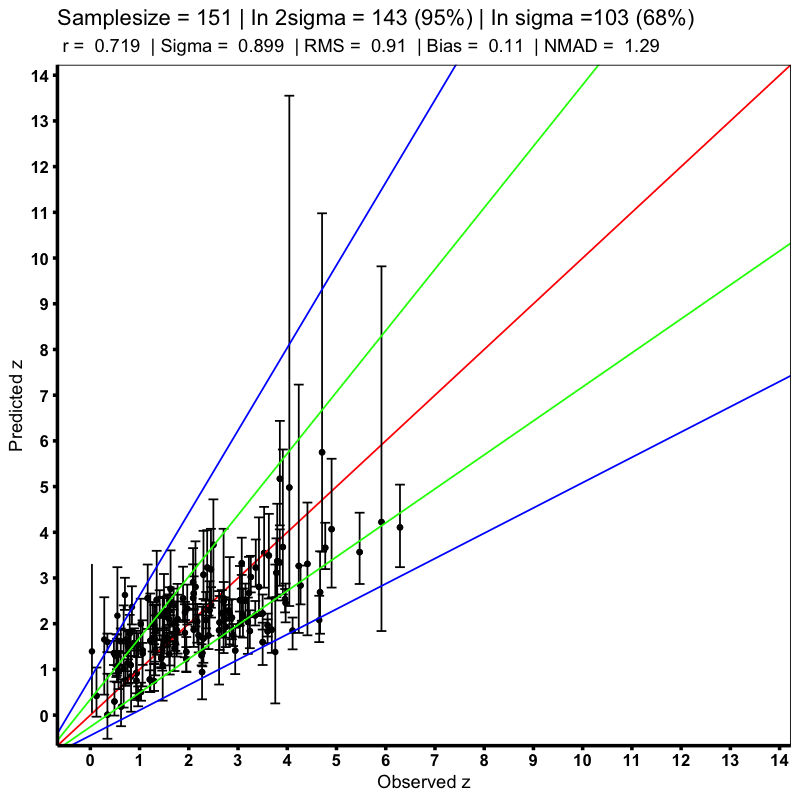}
    
     \includegraphics[scale=0.27]{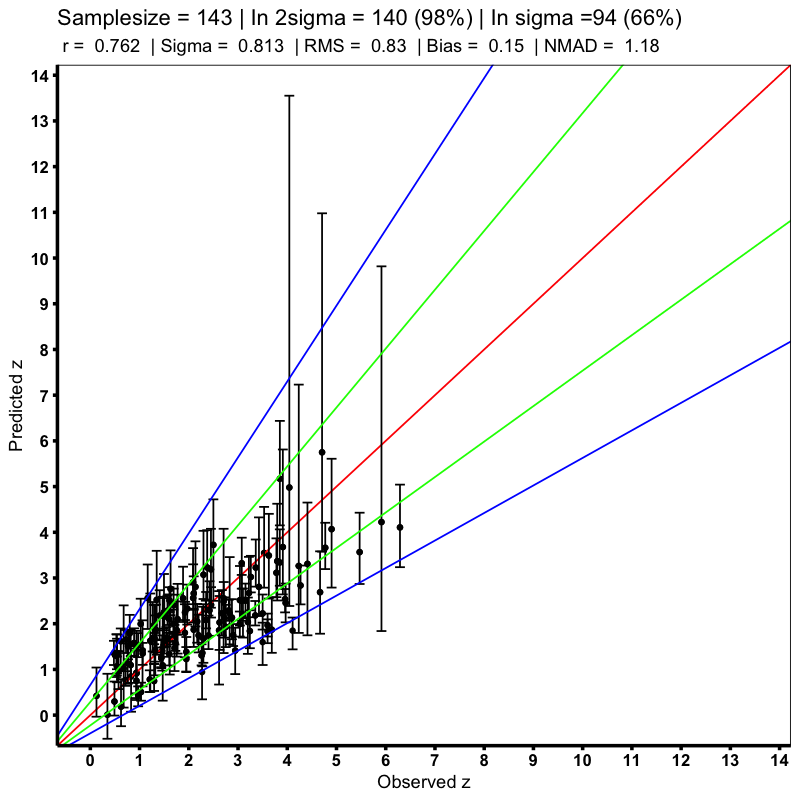}
     
     \includegraphics[scale=0.27]{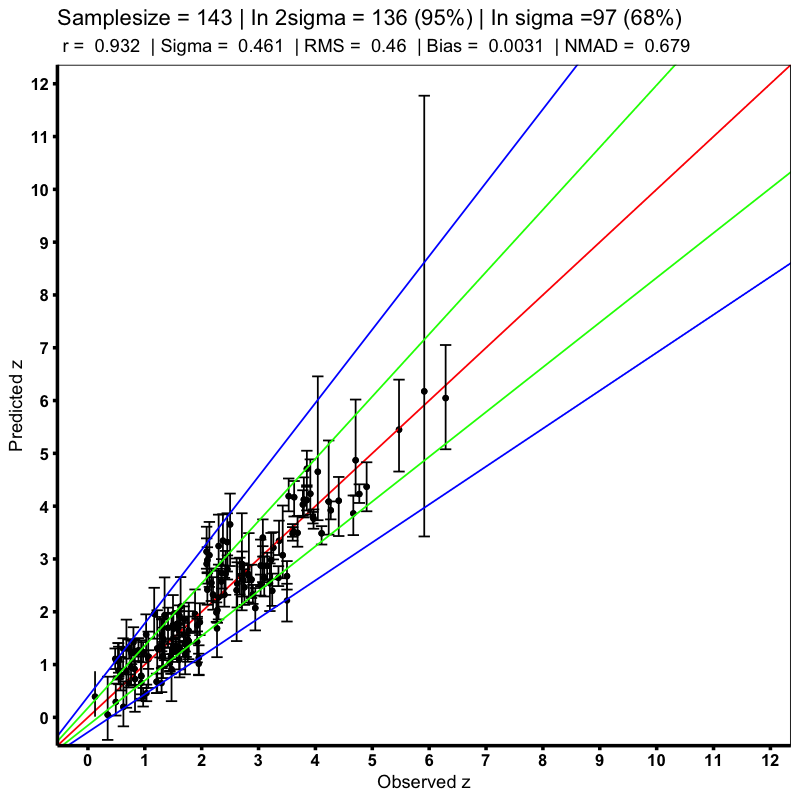}
     
    \caption{The scatter plot between $z_{obs}$ and $z_{pred}$. Upper panel: Predictions before removal of catastrophic outliers and bias correction. Middle panel: Predictions after the removal of catastrophic outliers.
    Lower panel: Predictions after the removal catastrophic outliers and application of bias correction.}
    \label{resultszpredvszobs}
\end{figure}

{The error bars on $z_{pred}$ are generated by MCMC simulations with Gaussian distributions centered around the central value of the observed variables and as standard deviation their measurement uncertainties.
This procedure is repeated 100 times for each GRB in the 10fCV algorithm and allows us to obtain a redshift distribution whose minimum and maximum represent the error bars on $z_{pred}$ (see the errorbars in the Fig. \ref{resultszpredvszobs}).}

We obtain the $1\sigma$ and $2\sigma$ cones in Fig. \ref{resultszpredvszobs} between $z_{pred}$ and $z_{obs}$.
The 1$\sigma$ and 2$\sigma$ are defined as:
$$
1\sigma = 10^{\sigma^*}z + (10^{\sigma^*}-1)
$$
$$
2\sigma = 10^{2\sigma^*}z + (10^{2\sigma^*}-1)
$$
where $\sigma^*$ is the standard deviation in the $\log(z+1)$ scale.

It is at this stage that we remove additional outliers deemed to be 'catastrophic'. 
These catastrophic outliers are defined in \cite{jones2020tests} as the GRBs which $|{\Delta z}| > 2\sigma$. 
In our case, 8 GRBs are catastrophic outliers, and as shown in Fig. \ref{resultszpredvszobs}, they fall outside of the blue cone. Following this additional outlier removal procedure, our sample of 151 GRBs is reduced to 143 GRBs.
It is important to note that we do not retrain our models following the catastrophic outlier removal.

\subsubsection{Bias Correction}\label{bias}
Following the removal of catastrophic outliers, we perform bias correction on our predictions.
Bias is defined as the mean of the difference between predicted and observed values of the response variable. 
When training models on an imbalanced sample or a sample that has some level of discrimination against a set of random variable outcomes, the model's predictions can have significant bias. 
With an imbalanced sample, the model becomes more skilled at predicting the redshift range with more observations.
To correct for this bias, we use the Optimal Transport bias correction technique. 
This involves sorting in ascending order the predicted and observed values and fitting a linear model between them. 
The bias is corrected using the slope and intercept of this fit following this formula:

$$Y_{pred}=\beta_0+\beta_1 Y_{SL}$$

\noindent where $Y_{preds}$ is the corrected predictions, $Y_{SL}$ is the SuperLearner predictions, and $\beta_0$ and $\beta_1$ are the intercept and slope of the linear fits, respectively. 
Solving for the fitted $Y_{pred}$ values provides the bias-corrected redshift estimates. 
We apply this technique separately to four independent regions, namely, $z_{obs} < 2$, $2 < z_{obs} < 3.5$, $3.5 < z_{obs} < 5.0$, and $z_{obs} > 5.0$. 
This allows us to adapt the bias correction to different ranges of $z_{obs}$, providing more accurate results.


\subsubsection{Predicting the redshift of the Generalization Set}\label{genPred}
Given that our model has been fully validated using the 10fCV, we may now begin making predictions for our generalization set.
{However, before we use the generalization data, we must first perform data cleaning. 
This process is similar to the data cleaning of the training set.
However, the main difference is that we now remove GRBs outside of the parameter space of our training set, in addition to the cuts performed to ensure a representative set of the parameters. 
This results in a total of 67 GRBs (29.8\% decrease) being removed out of 221 GRBs.
Thus, our generalization set consists of 154 GRBs.
When predicting the generalization set, it is important to make sure that the GRBs are within the parameter space of the trained SuperLearner model.
Otherwise, the model will extrapolate the redshift predictions for those GRBs, leading to lower confidence in their accuracy.
}
With the best model obtained with the SuperLearner, we predict the redshift {of these 154 GRBs} using the same predictors we have used in the training set.
Since there is no observed redshift data for this dataset, we checked that our predicted redshifts come from the same parent population of the observed redshifts. 
We perform the Kolmogrov-Smirnov test \citep{Karson1968} and Anderson-Darling test \citep{Stephens1974} to verify if the two sets of data share the same underlying distribution.


\section{Results}\label{results}

\begin{figure}
    \includegraphics[scale=0.7]{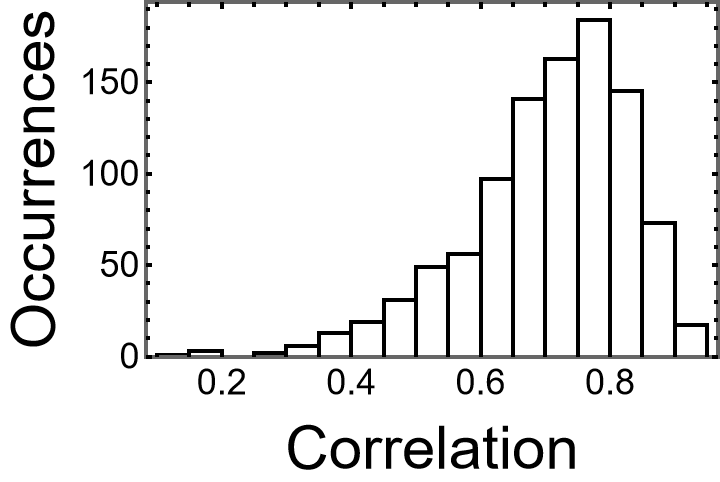}
    \includegraphics[scale=0.7]{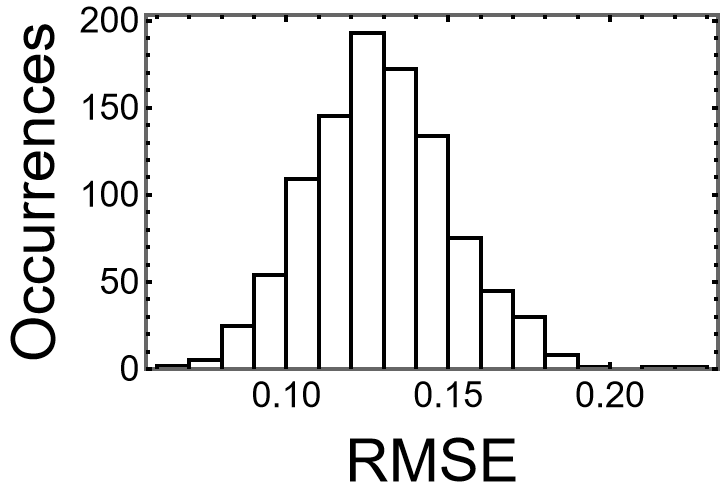} 
       \caption{Histograms of our chosen metrics on testing sets. Both correlation and RMSE are computed in the $\log(z+1)$ scale. 
       Left panel: correlation between predicted and observed redshift. 
       Right panel: RMSE of the predicted redshifts.}
    \label{fig:testsets}
\end{figure}

In this section, we present the results of our analysis, including the performance of GAM, GLM, and the SuperLearner ensemble. We also discuss the implications of our findings and the application of these models in estimating the redshift in the generalization set.

\subsection{Performance Metrics and Comparison}

The benchmark for evaluating the quality of our results revolves around minimizing the RMSE within the SuperLearner algorithm. 
To further evaluate the efficacy of our results, we also utilize the following metrics in conjunction with SuperLearner RMSE: the Pearson correlation coefficient ($r$) between $z_{obs}$ and $z_{pred}$, the normalized median absolute deviation (NMAD), and the bias defined as the mean of $z_{pred} - z_{obs}$.
{These metrics provide a reliable assessment of each model's ability to predict redshift values accurately.}



\subsection{SuperLearner Results}
\label{ensemble}




The Superlearner identified GAM and GLM as the best predicting models.
The ensemble generated with GAM {$(A_1 = 0.649)$}, exhibits the highest predictive capability, followed by GLM {($A_2 = 0.351$)}. 
In this context, $A_1$ and $A_2$ denote coefficients indicating the optimal model and reflecting the weighted average of multiple models.

The results obtained from SuperLearner are presented in Fig. \ref{resultszpredvszobs}.
We obtain a {$r$=0.719, RMSE=0.91, bias=0.11 and NMAD=1.29} between $z_{obs}$ and $z_{pred}$ including all GRBs.
The catastrophic outlier percentage is {5.3}\%.
Following the steps mentioned in Sec. \ref{errorbars} and Sec. \ref{bias}, the results after the catastrophic outliers are removed, as shown in the middle panel of Fig. \ref{resultszpredvszobs}.
Here, we obtain an improvement of the correlation which reach {r=0.762, RMSE=0.83 and bias=0.15}. 
The bias-corrected results are presented in the bottom panel of Fig. \ref{resultszpredvszobs}. 
Here, we see an improvement compared to the non-biased corrected results.
The improvement is visible in all the metrics: 
$r$={0.932 (29\% increase), RMSE=0.46 (49\% decrease) and NMAD=0.68 (47\% decrease)} between $z_{obs}$ and $z_{pred}$. 
The catastrophic outlier percentage also drops from 5.3\% to 4.9\%.
{For high-$z$ GRBs ($z_{obs}\ge$ 3), $r$=0.85, RMSE=0.5, and bias=0.17.}

\subsection{GAM and GLM Performance}

We begin by evaluating the predictive performance of the GAM and GLM models individually. 
We identified the most promising formulas for each model based on cross-validation metrics, specifically correlation, RMSE, and NMAD. 
We found that the selected formulas for both GAM and GLM models demonstrated substantial correlation ({0.65} and {0.66}, respectively) in $\log(z+1)$ scale and exhibited relatively low RMSE (0.99 and 1.20) and NMAD ({1.36} and {1.34}) values when performed on 100 10fCV. 
The GAM and the GLM formulae that perform the best in terms of correlation are quoted in Eq. \ref{eq:GAM} and Eq. \ref{RLM}, respectively. 

Results of SuperLearner on the test sets during 10fCV are presented in Fig. \ref{fig:testsets}, which shows the distribution of $r$ (left panel) and the RMSE (right panel) of the 100 runs of the nested-fold 10fCV procedure.
Our results show that the accuracy of the prediction is quite stable with a correlation coefficient between the $z_{pred}$ vs. $z_{obs}$ which peaks at {0.75} and an RMSE which peaks at 0.13 for the majority of partitions in the training sets and test sets.
However, for a small number of partitions, we observe a small correlation coefficient and a higher RMSE. 
This is indeed natural due to the large heterogeneity of the data and the relatively small sample size. 

\subsection{Predicting the Generalized Data Set}\label{generalizationsetprediction}
 
The distributions of predicted redshift of the generalization set and $z_{obs}$ of the training set are presented in Fig. \ref{fig:final_model_hist} with dashed and solid bars, respectively. 
The generalization set distribution has been obtained using the model after the optimal transport bias correction.
We have checked the Anderson Darling for these two distributions, and the hypothesis that they are drawn by the same parent population is rejected. 
We further investigated the reason why the two distributions are not compatible with each other and we investigated the distribution of each variable that comes into play. 
We observed that the distributions of two variables ($\log(T_{90}$) and $\log(NH)$) are also not drawn by the same parent population according to the Anderson Darling Test. 
Thus, possibly this is the cause of this discrepancy. 
In addition, in our initial sample we have removed the IS GRBs, but since the IS GRBs need to have a redshift to be defined, we do not have the possibility to highlight them. 
Another problem is that the classification of the SGRBs with extended emission for GRBs without redshift is not often reported in the literature, and thus, it is hard to classify GRBs appropriately in the generalization set.
{In order to further investigate this issue we added the IS GRBs to the sample and repeated the same procedure described above.
We note that the best GAM and the best GLM are the same as the ones detailed in the previous sections. 
The number of IS GRBs is 11. This means that our training sample has been increased by 6\% compared to the total sample of the training and the test set. 
This increase of the training sample has not changed the formula for the prediction telling us indeed that our prediction is stable at the increase of the sample at least at the level of 6\%.}

\begin{figure}[!h]
    \centering
    \includegraphics[scale=0.3]{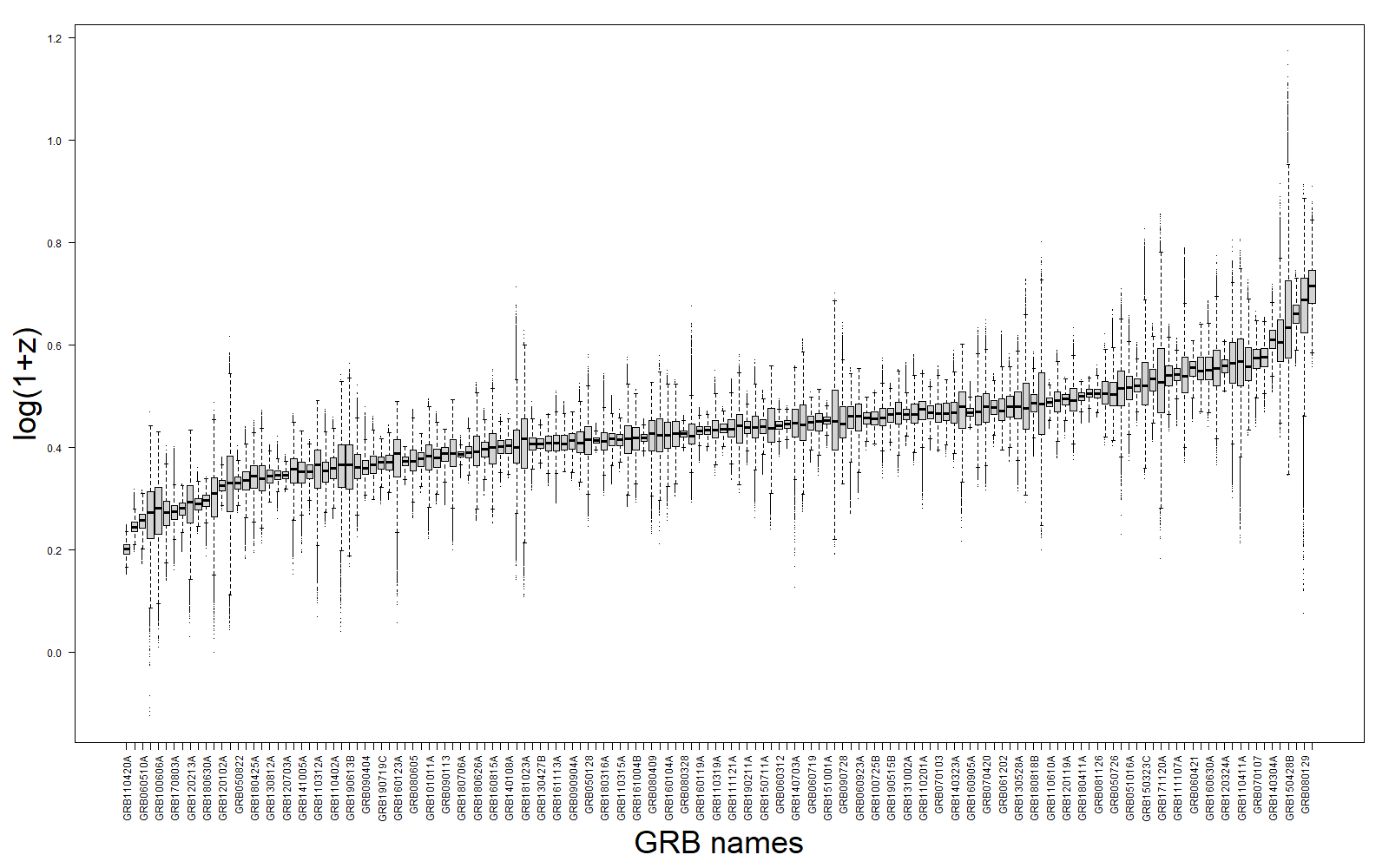}
    \caption{The box plot for each GRB is shown for the generalization set. The error bars are shown for each prediction.}
    \label{fig:boxplot}
\end{figure}

\begin{figure}[!h]
    \centering
    \includegraphics[scale=1]{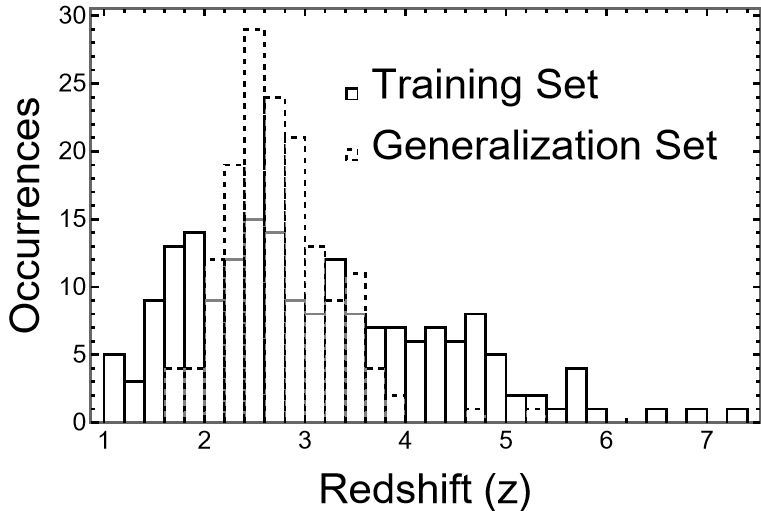}
    \caption{Histogram comparing the distributions of the training set $z_{obs}$ and the redshift predictions of the generalization set.}
    \label{fig:final_model_hist}
\end{figure}

\begin{figure}[b]
    \centering
    \includegraphics[width=0.32\textwidth]{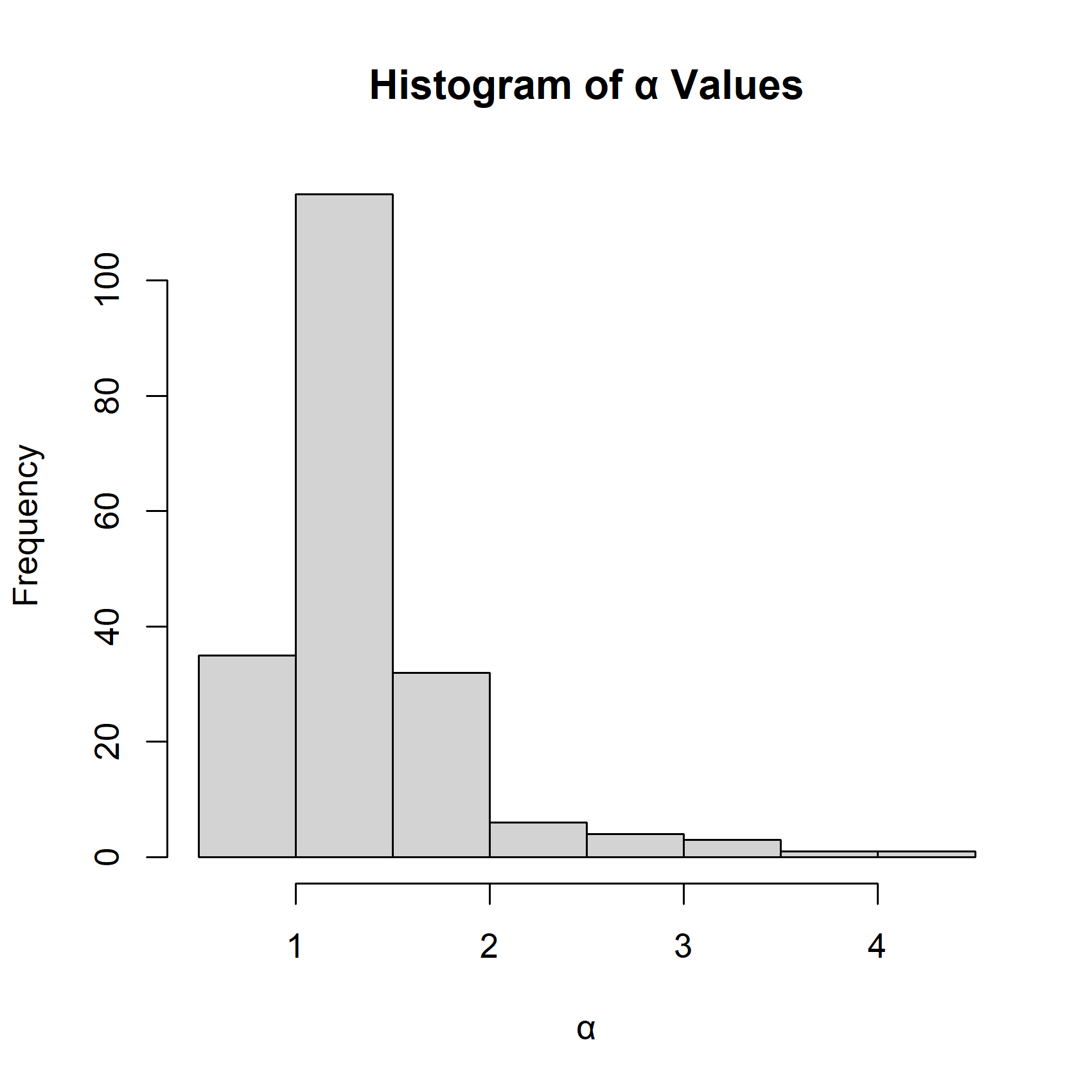}
    \includegraphics[width=0.32\textwidth]{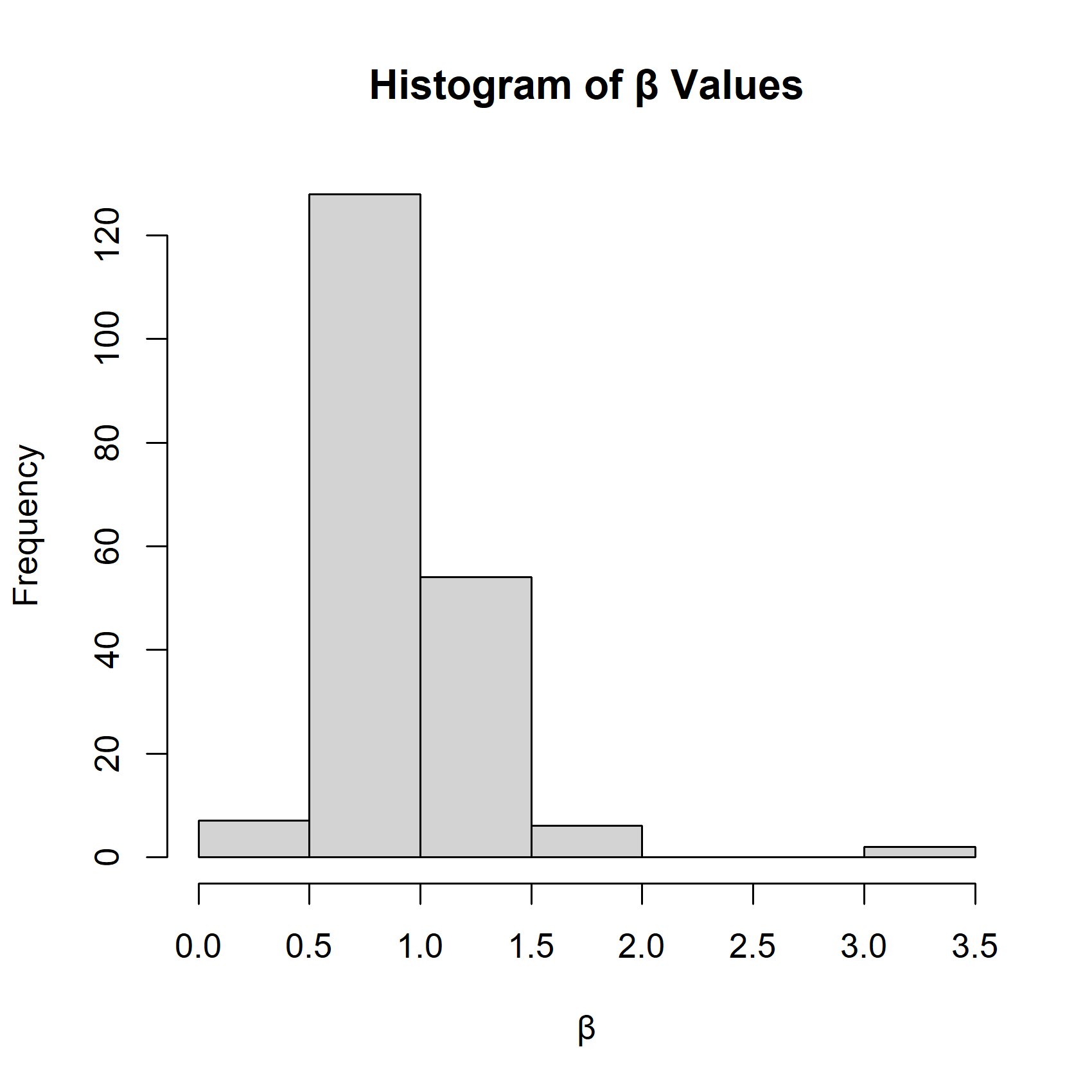}
    \includegraphics[width=0.32\textwidth]{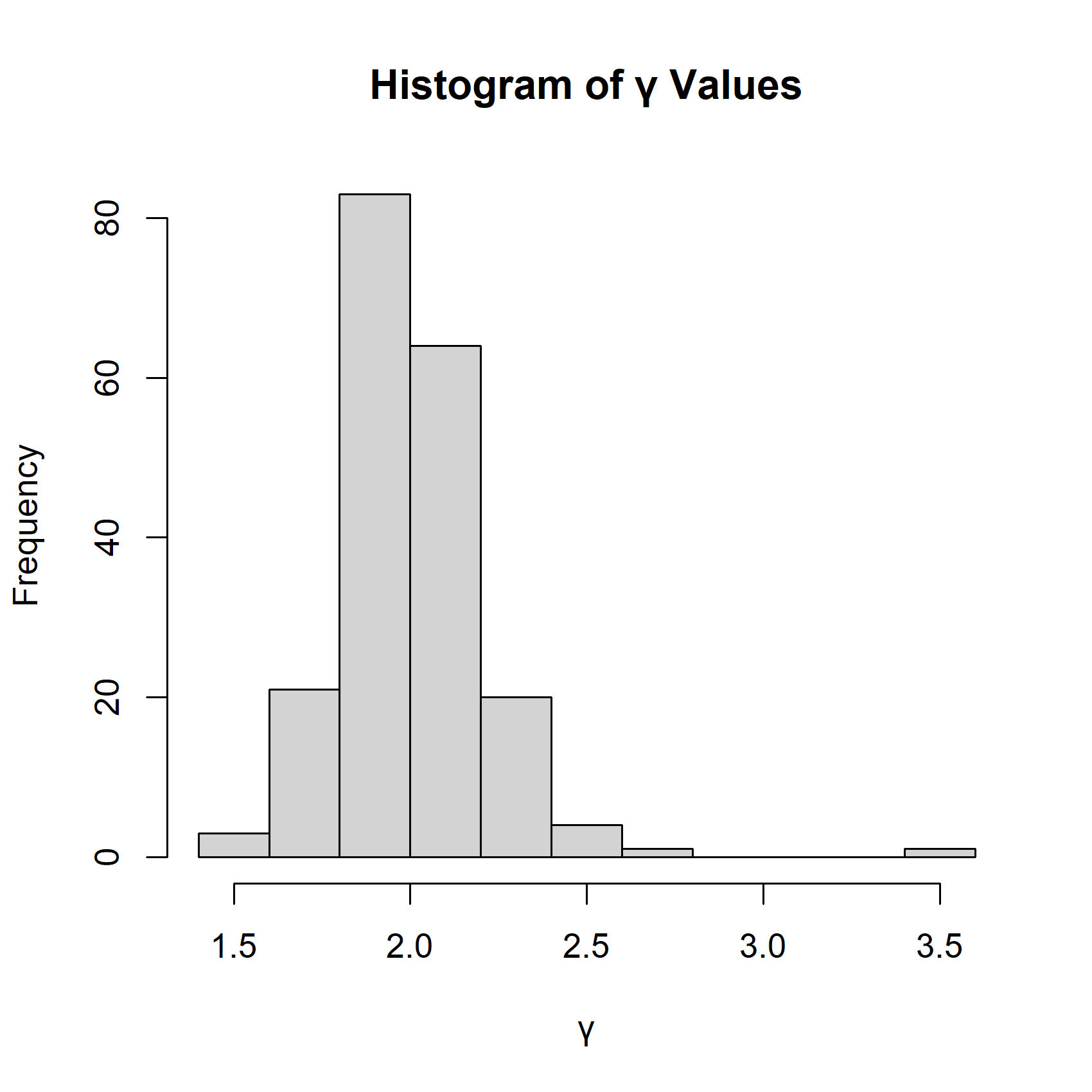}
    \caption{Histogram distributions for $\alpha$, $\beta$ and $\gamma$, showing the distribution before the outlier removal.}
    \label{fig:outlierhist}
\end{figure}

We have computed both the prediction errors and the error bars on the predictions following the approach detailed in Sec. \ref{errorbars}.
The contribution of these error bars are shown in the boxplot presented in Fig. \ref{fig:boxplot}.
It is clear for some GRBs the error measurements are large, but overall we can assess that our method is reliable to infer the redshift for GRBs for which the redshift is unknown.
{We would also like to clarify that because of the size of the image, not all the names of the GRBs are present. 
We also cut GRBs that have a prediction error greater than the redshift predicted by our model. 
This results in GRB090518, GRB150817A, and GRB190604B being removed from the dataset, leaving us with 151 GRBs of the generalization set in Fig. \ref{fig:boxplot}.}

\subsubsection{Comparative Results}
Prior to our investigation, several groups have used linear or non-linear relations between relevant GRB parameters.
However, the redshift inferred with these methods has not yet led to accurate measurements
\citep{Reichart2001, yonetoku2004,Atteia2005}.
Even when the inferred redshift uncertainty is small (5\%), these measurements are provided only for a few cases \citep{Guiriec2016}. 
When the redshift is inferred from the correlation between the peak in the $\nu F_{\nu}$ spectrum, $E_{peak}$, and the energy emitted isotropically \citep{amati2002} during the prompt emission, $r$ obtained between $z_{obs}$ and $z_{pred}$ is $0.67$ \citep{Atteia2003}. 
These $z_{pred}$ estimates, tested on 17 GRBs, are accurate only by a factor of 2 \citep{amati2006}.
When the redshift is inferred from the bi-dimensional X-ray Dainotti correlation between $F_a$ and $T_a/(1+z)$, only 28\% of cases have small error bars, namely $(z_{pred}-z_{obs})/z_{obs} < 1$ \citep{Dainotti2011}. All these attempts are parametric.

Compared to \cite{ukwatta2016machine}, parameters like $T_{90}$, Photon Index, Fluence, and NH remain the same.
Thus, it is the plateau variables, used here for the first time, that enhance our results.
\cite{ukwatta2016machine} employed Random Forest to estimate redshifts using prompt parameters, obtaining a correlation of 0.57 between $z_{pred}$ and $z_{obs}$.
{Comparing our results we obtain a 38\% improvement in correlation for the non-bias corrected results (top panel of Fig. \ref{resultszpredvszobs}) and 63\% improvement in correlation for the bias-corrected results (bottom panel of Fig. \ref{resultszpredvszobs}.}
Another problem, hardly explored in the literature (besides our study), is to account for measurement errors of GRB variables used to train ML models. 
Further, comparing our results with \cite{racz2017new}, who achieved a $r$=0.67, we see a 22\% increase in our correlation in the $\log_{10}(z+1)$ scale when we apply the bias correction.
In addition, our methodology is more complete than this work, since we use the LASSO feature selection, the M-estimator, the nested 100 10fCV, the Superlearner and the bias correction. 
Furthermore, we performed an extensive search by minimizing the RMSE and maximizing the correlations. 
All these steps have not been performed in the mentioned proceeding.

\section{Summary, Discussion and Conclusion}\label{discussion}
In this paper, we have developed a methodology to predict the redshift of GRBs from the Swift catalogue using prompt, plateau and afterglow parameters derived from the BAT+XRT observations.
The steps we have followed are:
\begin{enumerate}
    \item Cleaning and imputing the missing variables with MICE;
    \item Removing the outliers using an M-estimator;
    \item Applying the LASSO method to select the most predictive feature to reduce the number of predictors given the small data sample;
     \item Using the SuperLearner to perform a nested 10fCV method on 25 different models to determine which models would be the most successful.
     \item We determine that our best ML models for the ensemble are GAM and GLM;
     \item Optimizing each model by performing an extensive search aimed to maximize the Pearson Correlation coefficient and minimize RMSE on a test data;
     \item Creating our final ensemble with Eq. \ref{eq:GAM} for GAM and Eq. \ref{RLM} for the GLM. 
     \item Performing a 4-way bias correction on our training set predictions.
    \item Predicting the redshifts for the generalization set, including their prediction errors and the estimated observational uncertainties on the predicted redshifts. 
    This is achieved using MCMC simulations based on the uncertainties on the variables (see Sec. \ref{generalizationsetprediction}).
    
\end{enumerate} 

In comparison to other attempts to infer GRB redshift, our ensemble achieves an increase of 63\% and 38\% in the correlation between predicted and observed redshift reaching $r=0.93$, compared to other works in which only random forest or gradient boosting alone were used. \cite{Ukwatta2016} found $r=0.57$ with random forest, while \cite{Rcz2017} found $r=0.67$ both with random forest and gradient boosting. 
The main difference, besides our enhanced prediction and the fact that we use a more complete methodology which has been detailed in Sec. \ref{methodology}, is the use of the plateau properties.

In this work we highlight that the use of ML techniques for redshift prediction in GRBs offers several advantages over the commonly used parametric methods which employs only the use of bidimensional relations. 
Our individual GAM and GLM models exhibit strong predictive capabilities and demonstrate how to surpass the accuracy of existing approaches. 
In general, our study has shown that the use of parametric and semi-parametric methods brings an enhanced performance compared to the fully non-parametric approaches, like random forest, which are more prone to over-fitting compared to our methods.
In addition, the advantage for the parametric and semi-parametric models are more interpretable than the non-parametric models.
We have also shown that improved performance are actually happening with a reduced set of variables which contain first-order terms exclusively. 

Further, our study is a proof of concept for utilizing GRBs as standard candles in cosmology with the newly estimated redshifts from the generalization set. 
We have indeed increased the sample size of GRBs with known redshift by adding 154 GRBs (we have 423 GRBs with redshift so far) by 50\% of the total sample of GRBs with redshift.
With this new number we can use the Dainotti relation to serve as reliable cosmological tool. 
If we consider the increase of the estimates of GRBs with X-ray plateaus, then the increase is 94\%.

Looking towards the future, this is a preparatory work which will allows us to find values of $\Omega_{M}$ with a similar precision as \cite{Conley2011}.
As discussed in \cite{Dainotti2022c}, we would require 789 GRBs with X-ray plateaus to reach such a precision if we use GRBs which possess X-ray plateaus (see Table 9 in \cite{Dainotti2022b}). 
Since in the current analysis we obtain 154 GRBs with unknown redshift and X-ray plateaus, we currently have a sample of 222+154=376 GRBs with redshifts both known and inferred. 
Since we have a yearly rate of 15 GRBs with X-ray plateaus observed with redshift and 15 observed without redshift, our sample can be incremented from August 2019 until December 2023 by an additional 124 GRBs. 
This means that once we analyze the available GRBs with plateaus we will have more than 500 GRBs with known and inferred redshift. 
This will leave us only to wait for 789-500=289 GRBs to be observed, which with a rate of 30 GRBs with X-ray plateaus per year, will be reached in roughly 10 years. 
However, if we apply the lightcurve reconstruction analysis \cite{Dainotti2023a} we will have an higher precision on the plateau parameters (on average 37.5\%) which will allow us to need less GRBs, namely 37.2\% less of the initial total sample as detailed in \cite{Dainotti2022c}.
Therefore, from the total sample of 789 we should remove 37.2\% of it which is 293 GRBs. Thus, we would need 496, out of these we have already 489. 
This leaves us to wait only half a year to then reach the precision of \cite{Conley2011}.  
Thus, with the aid of both ML and lightcurve reconstruction, we can reach a higher precision in cosmology than the one obtained by \cite{Betoule2014}. 
For this analysis, we would need 987 GRBs, as detailed in \cite{Dainotti2022c}, so 987-489=498 GRBs more that can be observed in roughly 18 years. 
However, if we consider the uncertainties of the parameters of the plateaus divided by half the number of GRBs needed will be almost half of this sample, so the precision of \cite{Betoule2014} can be reached only in 9 years.

In conclusion, this work successfully predicts the redshifts of GRBs with X-ray plateau emission and introduces a pipeline of data processing techniques for obtaining reproducible and reliable results.
Through preprocessing, imputation, model selection, and performance evaluation, we predict redshifts for 154 LGRBs. 

Our results are a proof-of-concept showing the potential of ML-based methods to enhance the field of astrophysics and cosmology. 
By expanding the dataset of GRBs with known redshifts we can successfully tackle population studies such as the more accurate estimate of the luminosity function and density rate evolution, enabling a deeper understanding of the high-$z$ universe and its evolution. 
As further data becomes available and ML techniques continue to evolve, the accuracy and precision of redshift predictions for GRBs are likely to improve, offering exciting prospects for future cosmological research.

\section{Appendix}

{Here we present the distributions of $\alpha$, $\beta$, and $\gamma$ before the cuts mentioned in Sec. \ref{sec:datacleaning}. 
Based on the histograms, $\alpha$, $\beta$, and $\gamma$ values $>3$ are clearly outliers, as they belong to the tail of their respective distributions.
Because we wish to have the parameter space of the training data correspond as much as possible to the overall trend of the distributions for all the variables at play, thus, we impute them with MICE before performing the prediction analysis.}

\subsection{Results without removing outliers}
\begin{figure}
    \centering
    \includegraphics[scale=0.5]{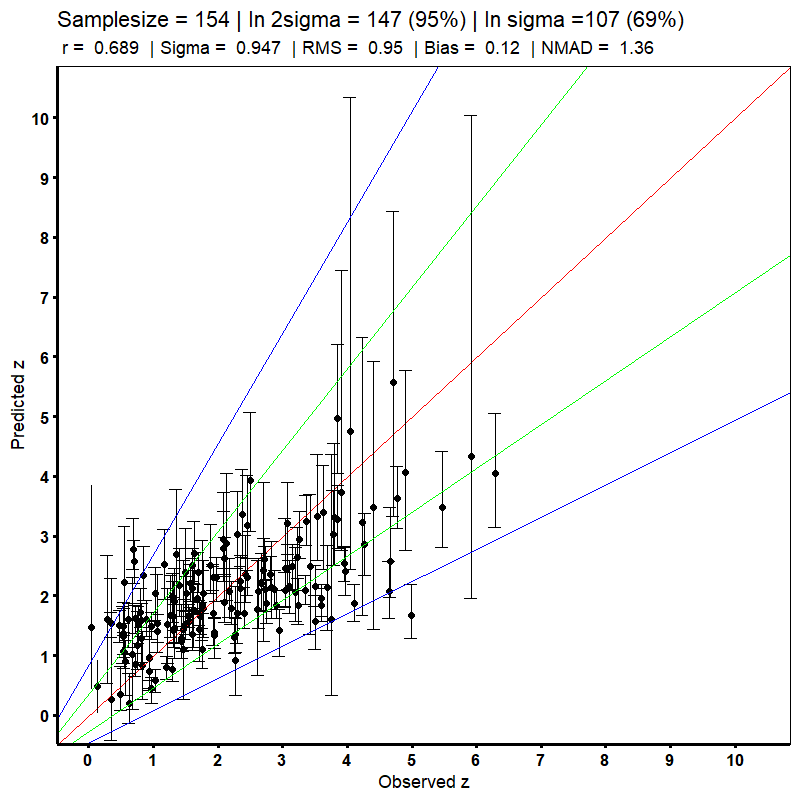}

    \caption{The scatter plot between $z_{obs}$ and $z_{pred}$. Predictions from the SuperLearner after 100 10fCV without using M-estimation outlier removal. 
    }
    \label{new_results}
\end{figure}

{Here we present the results obtained when we do not eliminate the M-estimator outliers.
Namely, here the training set is larger by 2\%, from 152 to 155.
We applied the exact same methodology described previously.
Below we present the 10fCV correlation plots for the SuperLearner. 
As can be observed in the Fig. \ref{new_results}, we see a 4\% reduction in correlation, a 5.5\% increase in RMSE, and a 0.73\% increase in the MAD.}

{Thus, these results conclusively show that removing the outliers with M-estimator does indeed improve the performance of our ML models.
However, given the small decrease in correlation and the small increase in RMSE we can observe a stable prediction in terms of these metrics.
}

\section{Acknowledgments}\label{sec:acknowledgments}

We acknowledge Artem Poliszczuk for the discussion about the extensive search. 
We would like to acknowledge the United States Department of Energy for its sponsorship of the Summer Undergraduate Laboratory Internship program which supported this work.
We acknowledge funding from Swift GI Cycle XIX, grant number 22-SWIFT22-0032.
This work was also supported by the Polish National Science Centre grant UMO-2018/30/M/ST9/00757 and by Polish Ministry of Science and Higher Education grant DIR/WK/2018/12.

\bibliographystyle{aasjournal}
\bibliography{bibliography}

\end{document}